\documentclass{aa}  
\usepackage{graphicx}
\usepackage{txfonts}
\usepackage[]{hyperref}
\usepackage{cleveref}
\usepackage{epstopdf}
\usepackage{todonotes}
\usepackage{multirow}
\newdimen\mylength
\newcommand{\hmsun}{h^{-1}{\rm M}_\odot}
\newcommand{\hmpc}{h^{-1}{\rm Mpc}}

\begin{document} 
  
   \title{How galaxies populate haloes in very low-density environments?}
   \subtitle{An analysis of the Halo Occupation Distribution in cosmic voids}

   \author{Ignacio. G. Alfaro \thanks{E-mail:german.alfaro@unc.edu.ar}, Facundo Rodriguez, Andr\'es N. Ruiz \& Diego Garcia Lambas}
   
   \authorrunning{I. G. Alfaro et al.}
   
   \institute{Instituto de Astronomía Teórica y Experimental, CONICET-UNC, Laprida 854, X5000BGR, C\'ordoba, Argentina \\ Observatorio Astron\'omico de C\'ordoba, UNC, Laprida 854, X5000BGR, C\'ordoba, Argentina.}

   \date{\today}

  \abstract
   {Evidence shows properties of dark matter haloes may vary with large-scale environment. By studying the halo occupation distribution in cosmic voids it is possible to obtain useful information that can shed light on the subject. The history of the formation of the haloes and galaxies residing in these regions is likely to differ from the global behaviour given their extreme environment.}
   {Our goal is to characterize the halo occupation distribution in the interior of cosmic voids and compare with the general results to unveil the way galaxies populate haloes in simulated galaxy catalogues.}
   {We use two public access simulated galaxy catalogues constructed with different methods: a semi-analytical model and a hydrodynamic simulation. In both, we identify cosmic voids and we measure the halo occupation distribution inside these regions for different absolute magnitude thresholds. We compare these determinations with the overall results and we study the dependence of different characteristics of the voids. Also, we analyze the stellar content and the formation time of the haloes inside voids and confront the general halo population results.}
   {Inside the voids, we find a significantly different halo occupation distribution with respect to the general results. This is present in all absolute magnitude ranges explored. We obtain no signs of variation related to void characteristics indicating that the effects depend only on the density of the large-scale environment. Additionally, we find that the stellar mass content also differs within voids, which host haloes with less massive central galaxies ($\sim 10\%$) as well as satellites with significantly lower stellar mass content ($\sim30\%$). Finally, we find a slight difference between the formation times of the haloes which are younger in voids than the average population. These characteristics indicate that haloes populating voids have had a different formation history, inducing significant changes on the halo occupation distribution.}
  {}

   \maketitle
  
%

\section{Introduction}

    The current paradigm for structure formation in the Universe assumes galaxies forming by baryon condensation within the potential wells defined by the collisionless collapse of dark matter haloes \citep{White1978}. However, the diversity of astrophysical mechanisms involved in the process of galaxy formation and evolution does not allow to determine unambiguously how galaxies occupy haloes. Thus, understanding the links between galaxies and the dark matter haloes in which they reside is one of the keys to comprehending the formation and evolution of large structures.

    Several works show  that the Halo Occupation Distribution (HOD) is a powerful tool to connect galaxies and dark matter haloes. HOD describes the probability distribution $P(N|M_{\rm halo})$ that a virialized halo of mass $M_{\rm halo}$ contains $N$ galaxies with some specified characteristic \citep[e.g.][]{Jing1998,Ma2000, Peacock2000,Seljak2000,Scoccimarro2001, Berlind2002,Cooray2002,Berlind2003, Zheng2005, Yang2007, Rodriguez2015}. Moreover, many authors have proposed the use of the HOD to constrain models of galaxy formation and evolution \citep[e.g.][]{Benson2000, Berlind2003, Kravtsov2004,Zentner2005,Zehavi2011} and to cosmological models  \citep[e.g.][]{Vandenbosch2003,Zheng2007}. 
    
    Generally, the HOD approach assumes that the halo population depends only on their mass, a first approximation that has been ample analyzed. Some examples of this, are the works of \cite{Pujol2014} who found that HOD cannot reconstruct the galaxy bias for low mass haloes in several semi-analytic models, and \cite{Pujol2017}  semi-analytic models analysis of the influence of local density. In the later, the authors point out HOD as a better predictor of galaxy bias than halo mass, while on the other hand,  \cite{Berlind2003} study of the environmental variations of the HOD in hydrodynamical cosmological simulations shows no significant dependence. However,  \cite{Zehavi2018},  \cite{Artale2018} and \cite{bose_hod_2019} explored the dependence of halo occupation on the large scale environment for some semi-analytic and hydrodynamical simulations, finding some relevant signs of correlation.

    As it is well known, the cosmic web that forms the large-scale structure of the Universe contains regions with large mass density fluctuations with respect to the mean background density, with cosmic voids corresponding to the lowest density regions. There are several precise definitions of cosmic voids, where the differences of these definitions account for a diversity of derived void properties such as their topology or the inner mass content \citep[]{Colberg2008,Cautun2018}. Nevertheless, all the definitions agree that these regions comprise most of the volume of the Universe, but only contain a small fraction of the galaxy populations \citep{Pan2012}. 
    
    The extremely low densities plus the expansion of the void turn gravitational interactions between galaxies less frequent, a fact that affects the growth and fate of structure. These characteristics make cosmic voids ideal regions to study aspects of the formation and evolution of galaxies unlikely to be observed elsewhere. The influence of these extremely low-density environments on their member galaxies may be reflected in different dynamical and astrophysics properties compared to galaxies populating higher density regions. Void galaxies tend to be blue, faint and of late-type morphologies \citep[]{Rojas2004,Hoyle2005,Patiri2006,Ceccarelli2008,Hoyle2012}, generally exhibiting young stellar populations and intense star formation activity \citep{Rojas2005}. Thus, void galaxies are expected to have a significantly different dynamical and astrophysical evolution. In this context, the possible dependence of HOD on the environment should be most clearly seen inside cosmic voids.  
    
     In previous work, \cite{ruiz_into_2019} found a significantly low amplitude of the galaxy-galaxy correlation function for galaxy samples inside cosmic voids. As analysed by \cite{Cooray2002}, the galaxy power spectrum can be expressed in terms of the HOD. Thus, given that the correlation function is straightforwardly associated with the power spectrum \citep{Peebles1980}, HOD is also expected to differ when measured inside the cosmic voids. In this work, we use cosmic void definition following \cite{ruiz_into_2019} that considers spherical regions where the integrated density contrast doe not exceed a given threshold value $\Delta_{lim}$. In the literature, see for instance \citet{ruiz_void_2015} and references therein, this parameter varies between $-0.8$ and $-0.9$, so that inside void boundaries there is at most $20 \% $ to $10 \%$ the mean density of the Universe.
    
    This paper is organized as follows. In Sec. \ref{sec2} we describe the simulated galaxy catalogues obtained from both, a semi-analytic model , and a hydrodynamical simulation. We also describe the algorithms to identify cosmic voids and the void catalogues obtained. In Sec. \ref{sec3} we describe the methodology used to determine the HOD inside cosmic voids. In Sec. \ref{sec4} we present the results of the HODs measurements inside the voids in both catalogues. In this section, we also explore the dependence of the results on void properties such as $\Delta_{\rm lim}$, void size, and the surrounding void environment. In Sec. \ref{sec5} we study the halo stellar mass distribution as a function of the total dark matter halo mass. In Sec. \ref{sec6} we compare the halo formation time inside voids with the overall results. Finally, in Sec. \ref{sec7} we present our summary and conclusions. 

\section{Data}\label{sec2}

In this section we present the simulated galaxy catalogues used in this work. We also present a brief description of the void identification algorithm used, and the final void catalogues obtained. 

\subsection{Simulated galaxy catalogues}

We use two simulated galaxy catalogues, one based on a semi-analytic approach, and one extracted from a hydrodynamic simulation.

\subsubsection{The MDPL2-SAG galaxy catalogue}

The MDPL2-SAG\footnote{doi:10.17876/cosmosim/mdpl2/007} catalogue is part of the MultiDark-Galaxies catalogues \citep{knebe_multidark_2018}, publicly available at the CosmoSim\footnote{https://www.cosmosim.org} and Skies \& Universes\footnote{https://www.skiesanduniverses.org} databases.
This catalogue was constructed by applying the semi-analytic model of galaxy formation and evolution SAG (acronym for Semi-Analytic Galaxies) to the dark matter haloes of the MDPL2 cosmological simulation. 

The SAG model \citep{cora_sag_2018} includes all the main physical processes involved in the formation and evolution of galaxies, such as star formation, supernova feedback, radiative cooling of the hot gas, chemical enrichment of gas and galaxies, growth of supermassive black holes, feedback by AGN, starbursts via disc instabilities and galaxy mergers.
For readers interested in a complete description of the several physical processes present in the SAG model and the details of its implementation, we refer to \citet{cora_sag_2006}, \citet{lagos_sag_2008}, \citet{tecce_sag_2010}, \citet{ruiz_sag_2015}, \citet{gargiulo_sag_2015} and \citet{cora_sag_2018}.  

The MDPL2 simulation \citep{riebe_multidark_2013,klypin_multidark_2016}, which is also available at the CosmoSim database, counts with 3840$^3$ dark 
matter particles in a comoving box of 1000$\hmpc$ on a side, which translates 
in a particle mass resolution of $1.51\times 10^9 \hmsun$.
The cosmological parameters adopted correspond to a flat $\Lambda$CDM scenario 
consistent with Planck results \citep{planck_2014}: $\Omega_{\rm m}=0.307$, 
$\Omega_{\rm b} = 0.048$, $\sigma_8 = 0.823$, $h=0.678$ and $n=0.96$.

The haloes and subhaloes used to populate the simulation with galaxies were identified with the Rockstar Halo Finder \citep{behroozi_rockstar_2013} and the merger trees were constructed with ConsistentTrees \citep{behroozi_trees_2013}. 

From the complete MDPL2-SAG catalogue at $z=0$, we select all galaxies with absolute magnitudes in the $r$-band $M_r - 5\log_{10}(h) \le -16$, stellar masses $M_{\star} \ge 5\times 10^8 \hmsun$ and host haloes with masses $M_{200c} \ge 5\times 10^{10} \hmsun$, where $M_{200c}$ corresponds to the mass enclosed within overdensity of 200 times the critical density of the Universe. The final catalogue comprises 41986893 galaxies.

\subsubsection{The TNG300 galaxy catalogue}

The TNG300 simulation is part of the IllustrisTNG\footnote{http://www.tng-project.org/} suite of hydrodynamical simulations of galaxy formation in cosmological volumes \citep{marinacci_tng_2018,naiman_tng_2018,nelson_tng_2018,pillepich_tng_2018,springel_tng_2018}. 
Particulary, the TNG300 counts with $2500^3$ dark matter particles and $2500^3$ gas particles in a cubic comoving box of $205\hmpc$ on a side, which results in a dark matter and baryonic mass resolutions of $3.98\times 10^7 \hmsun$ and $7.44\times 10^6 \hmsun$, respectively. 
The cosmological model adopted is a flat $\Lambda$CDM with parameters also in agreement with Planck results \citep{planck_2016}: $\Omega_{\rm m} = 0.3089$, $\Omega_{\rm b} = 0.0486$, $\sigma_8 = 0.8159$, $h = 0.6774$ and $n=0.9667$.  
The simulation was evolved using the moving mesh code AREPO \citep{springel_arepo_2010} and includes not only all the relevant processes of galaxy formation but also a detailed magneto-hydrodynamical implementation (see \cite{weinberger_tng_2017} and \cite{pillepich_tng0_2018} for a complete description of the physical processes implemented).

We select all galaxies from the snapshot at $z=0$ with $M_r - 5\log_{10}(h) \le -15$ and host haloes with masses $M_{200c} \ge 10^{10} \hmsun$. The final catalogue counts with 657040 galaxies.

\subsection{Void identification and void catalogues}\label{2.2}

In both MDPL2-SAG and TNG300 galaxy catalogues, cosmic voids were identified using the algorithm presented in \citet{ruiz_void_2015}. 
Briefly, the identification starts estimating the density field via a Voronoi tessellation of the galaxy catalogues, which are used as tracers. 
For each Voronoi cell we can compute a density given by $\rho_{\rm cell} = 1/V_{\rm cell}$, being $V_{\rm cell}$ the volume of the cell, and define a density contrast as $\delta_{\rm cell}+1=\rho_{\rm cell}/\bar{\rho}$, where $\bar{\rho}$ is the mean density of tracers. 
All Voronoi cell which satisfies $\delta_{\rm cell} < -0.5$ is selected as a centre of an underdense region, and for those centres, we select as void candidates all-spherical volumes with an integrated density contrast $\Delta(R_{\rm void}) \le \Delta_{\rm lim}$, where $R_{\rm void}$ is the void radius and $\Delta_{\rm lim}$ is a density contrast threshold, usually chosen as $-0.8$ or $-0.9$ (which means 20\% or 10\% of the mean density of tracers, respectively).
For each of those void candidates, the computation of $\Delta$ is repeated several times in randomly displaced centres near the previous ones, accepting a new centre only if the new void radius is larger than the older one. 
This random walk procedure is performed in order to obtain void candidates centred the closest as possible in the true minimum of the density field.
The final step of the identification removes overlapping void candidates by keeping the largest voids which do not superpose with any other candidate. 

\begin{figure}
\begin{center}
\includegraphics[width=\columnwidth]{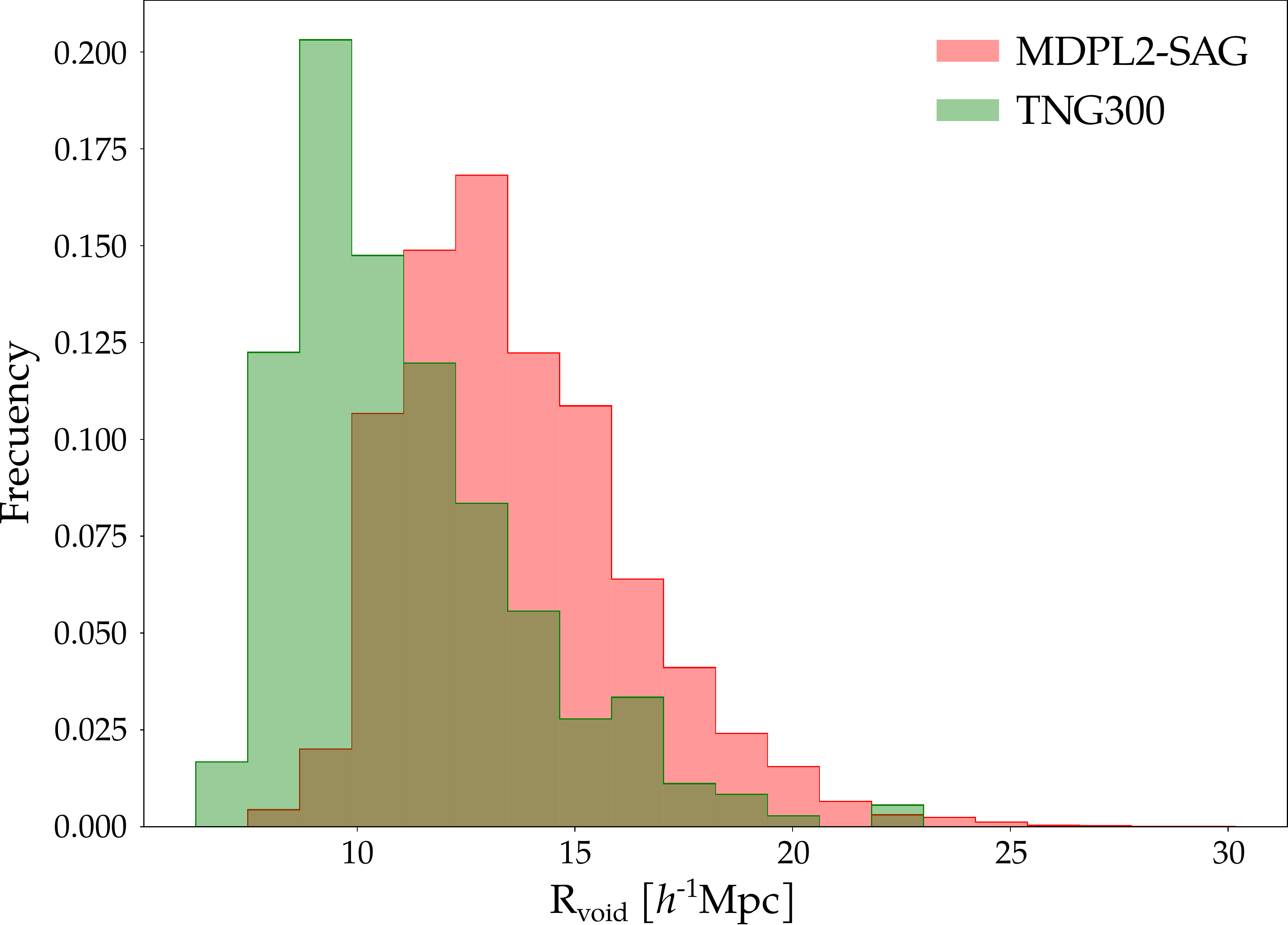}
\end{center}
\caption{\label{fig:RvoidHist}Radii distribution of cosmic voids identified in MDPL2-SAG (red) and TNG300 (green) catalogues using $\Delta_{\rm lim}=-0.9$.}
\end{figure}

In both catalogues used in this work, we select all galaxies with $M_r - 5\log_{10}(h) \le -20$ as tracers to identity voids, obtaining 12791 voids for MDPL2-SAG and 301 voids for TNG300.
In Fig. \ref{fig:RvoidHist} we show the radii distribution for both catalogues (MDPL2-SAG in red and TNG300 in green) and in Table \ref{Tabla} we show the relevant numbers of the void catalogues obtained.

It can be seen that TNG300 void radii are smaller than in MDPL2-SAG. This is due to the fact that TNG300 catalogue has a larger number density of tracers in this luminosity range than MDPL2-SAG catalogue (see Table \ref{Tabla}). 
To study the impact of the volume number density of galaxies and voids, in Appendix we have tested our results using samples of voids limited by size, in order to obtain the same number density of voids in both catalogues (Sec. \ref{sec:ap1}), and with void catalogues identified 
with equal number density of tracers (Sec. \ref{sec:ap2}).
Furthermore, the results found in this work (Sec. \ref{sec4}, \ref{sec5} and \ref{sec6}) remain unchanged.

Taken into account that observational works usually consider HOD as a function of a fixed absolute limiting magnitude, $M_{\rm lim}$ and aiming a future comparison with real data, we show here the results obtained with a fixed luminosity threshold.

To complete the description of the void samples of each catalogue, we present in Table \ref{Tabla} the number of voids in each sample and sub-sample considered.

\begin{center}   
\begin{table}
\small\addtolength{\tabcolsep}{-0pt}
\begin{tabular}{|c|c|c|}
\hline
                                             &  MDPL2-SAG   &    TNG300    \\
\hline
   $\overline{n} ~~ [10^{-3} h^3 \rm{Mpc}^{-3}]   $   &   $4.6543$   &    $12.3177$ \\ 
\hline
   Total number of voids                      &   $12791$    &    $301$     \\ 
\hline
   Number of galaxies in voids               &   $2756005$  &    $34351$   \\ 
\hline
   $R_{\rm void} < 10\hmpc$                  &   $381$      &    $130$     \\ 
\hline
   $10\hmpc \leq R_{\rm void} < 15\hmpc$     &   $8981$     &    $144$     \\ 
\hline
   $15\hmpc \leq R_{\rm void} < 20\hmpc$     &   $3121$     &    $24$      \\ 
\hline
   $20\hmpc \leq R_{\rm void}$               &   $308$      &    $3$       \\ 
\hline
   Median $R_{\rm void}$ ~~ [$\hmpc$]          &   $13.74$    &    $11.06$     \\ 
\hline
\end{tabular}
\caption{\label{Tabla} Properties of the comic voids identified in MDPL2-SAG and TNG300 catalogues. $\overline{n}$ represent the mean number density of tracers used to identify voids, i.e. the mean number density of galaxies with $M_r - 5\log_{10}(h) < -20$.}
\end{table}    
\end{center}

\section{Analysis of HOD in voids}\label{sec3}

In this section we describe the methodology adopted to measure the HOD and in particular, inside cosmic voids.

We study  the mean number of galaxies in haloes of a given mass, $\langle N | M_{\rm halo} \rangle$, being $M_{\rm halo} = M_{200c}$ for both catalogues. In order to compute this quantity, we use the available membership information of galaxies to associate them to their host dark matter haloes. In this case, the HOD is just obtained by binning in halo mass and computing the average number of galaxies in each bin.

To compute HOD inside cosmic voids we follow the same procedure described above but taking into account only those haloes which are completely enclosed within void boundaries, thus we remove from our samples all haloes which have galaxies beyond the void radius.

We have compared the overall HOD results with those computed using only haloes inside voids. In all cases, to compute the variance obtained in HOD calculations we use the jackknife technique. For this purpose, we separate the sample of haloes in 50 equal number sub-samples and we compute HOD variations when we do not consider each one of this sub-samples in the measurements. We also test the results using 10, 100, 150 and 1000 sub-samples in the jackknife procedure finding that, for 50 or more sub-samples the variance values stabilize.

In order to explore for possible HOD dependence on void parameters, we have explored the results obtained as a function of $\Delta_{\rm lim}$, void radius and void type classification according to their environment (see Sec. \ref{sec4}). 
We have also compared the stellar mass content of haloes residing inside voids with that derived for the total halo sample. This comparison provides further information to understand the environmental dependence of the relation between halo total stellar mass and number of galaxies.

\section{Results}\label{sec4}

\subsection{Dependence on $\Delta_{\rm lim}$}

   \begin{figure*}
   \begin{center}
   \includegraphics[width=\textwidth]{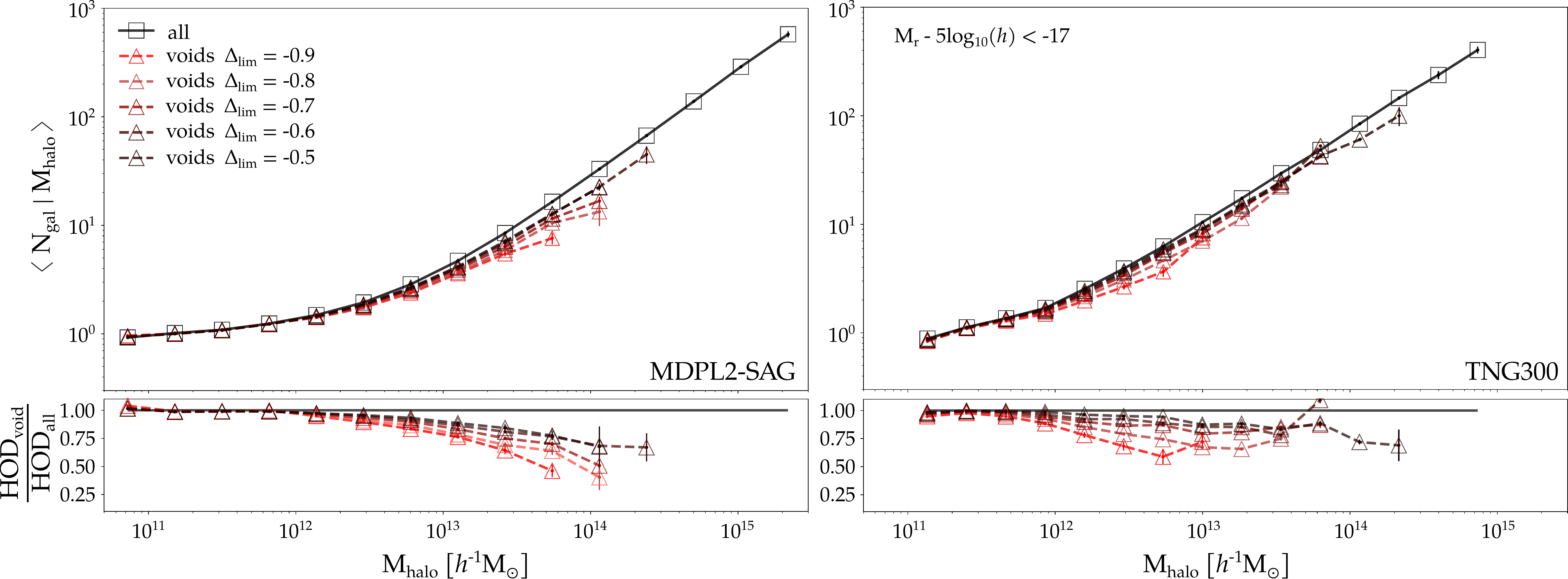}
   \end{center}
   \caption{\label{fig:HODvsDlim}HODs measured for galaxies with $M_{\rm r} - 5\log_{10}(h) \leq -17$ in MDPL2-SAG (left) and TNG300 (right) catalogues. The upper panels show the resulting HOD for the complete catalogues (black squares - solid lines) and for haloes inside cosmic voids (red triangles - dashed lines) identified with different values of $\Delta_{\rm lim}$, as indicated in the key. Bottom panels show the ratio between HOD inside voids and the HOD in the complete catalogues.  The error bars are computed using the standard jackknife procedure. }
   \end{figure*}

Cosmic voids are defined as spherical regions where the integrated overdensity is below a certain threshold value $\Delta_{\rm lim}$. This parameter is fundamental since it is the only free parameter in our identification algorithm. 
We identify voids with integrated density contrast  $\Delta_{\rm lim} = -0.5, -0.6, -0.7 , -0.8$ and $-0.9$. Within these structures, we measure the HOD in each void sample and compare the results with the global HOD obtained for the total haloes sample. 

In Fig. \ref{fig:HODvsDlim} we show the dependence of HOD with $\Delta_{\rm lim}$ in both MDPL2-SAG and TNG300 catalogues. 
The overall HOD is presented in black lines while red lines correspond to the HOD inside cosmic voids. 
As it can be seen, there is a clear dependence of the mean number of galaxies on the overdensity threshold used to identify the voids. This dependence is stronger in the case of MDPL2-SAG, however, TNG300 follows the same trend.  
By using less restricted $\Delta_{\rm lim}$ values, the HOD inside the voids becomes similar to that obtained in the complete catalogue. 
This dependence is in agreement with previous results presented by \citet{Zehavi2018}, \citet{Artale2018} and \citet{bose_hod_2019} who find evidence of occupancy variation when considering halo sub-samples taking into account their large-scale environment.
However, it is worth to mention some important differences between their approach and our analysis. 
These authors characterize the environment using spherical volumes \citep{Artale2018, bose_hod_2019} or a Gaussian smoothing \citep{Zehavi2018}, both with a fixed scale of 5$h^{-1}$Mpc, and determine the local density for each dark matter halo.
These criteria are substantially different than our spherical void definition, which involve large-scale underdensities with a fixed integrated density contrast, $\Delta_{\rm lim}$. 
A consequence of this definition is that, by construction, our voids have a lack of high mass haloes so that the internal void HOD is limited to a certain halo mass, as can be seen in Fig. \ref{fig:HODvsDlim} where the maximum halo mass that is achieved decreases with $\Delta_{\rm lim}$. For these reasons, given that the methodology and the halo mass range of HOD are different, a direct comparison between our results and those presented by \cite{Zehavi2018}, \cite{Artale2018} and \cite{bose_hod_2019} is not straightforward.
As a final remark, in Fig. \ref{fig:HODvsDlim} can be seen that beyond halo masses of $\sim 10^{12}\hmsun$, the environmental differences become significant, reaching factors as large as 2 for $\Delta_{\rm lim} = -0.9$. 

Since the aim of this work is to study the behaviour of HOD in very low-density environments, thus we decide to use voids identified with $\Delta_{\rm lim}=-0.9$. This selection is also justified by the fact that values of $\Delta_{\rm lim}=-0.8$ and $-0.9$ (which represent 20\% and 10\% of the mean density of tracers, respectively) are widely used in spherical void finder algorithms in the literature \citep[e.g.,][]{padilla_void_2005,ceccarelli_voids_2006, Colberg2008}.

\subsection{HOD inside voids}\label{4.2}

\begin{figure*}
\begin{center}
\includegraphics[width=0.85\textwidth]{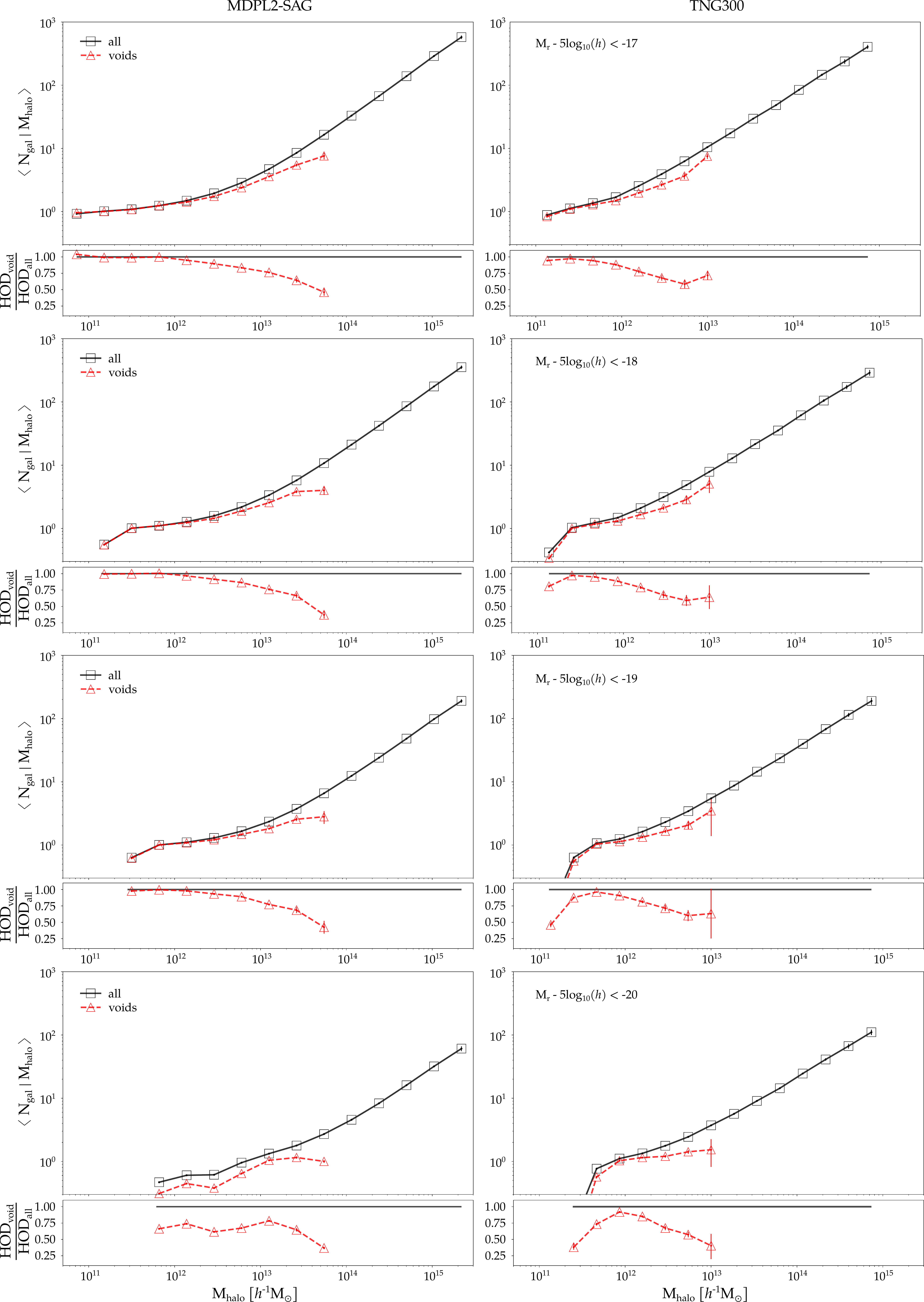}
\end{center}
\caption{\label{fig:HOD_voids}HOD in MDPL2-SAG and TNG300 catalogues for different luminosity thresholds. Left column shows the results for MDPL2-SAG and right column for TNG300 catalogue, for magnitude limits $M_r - 5\log(h)$ ranging from -17 to -20, from up to bottom. Black solid lines represent the overall HOD, meanwhile red dashed lines the HOD measured inside voids. For each magnitude bin and catalogue, the ratio between both HODs are showed at the bottom of each panel. The uncertainties are calculated by the standard jackknife procedure.}
\end{figure*}   

Once $\Delta_{\rm lim}$ value is fixed as $-0.9$, we compare the overall HOD with measurements inside voids for different $r-$band absolute magnitude thresholds, in order to analyze a possible dependence of the variation in the number of satellites in haloes inside voids with luminosity. 

In Fig. \ref{fig:HOD_voids} we compare the HOD inside voids and the overall results obtained on each catalogue for different absolute magnitude limits. 
The HOD inside voids is shown in red dashed lines and the overall HOD in black solid lines. The ratio between both is presented at the bottom of each panel.
As it can be seen, for all magnitude thresholds ($M_r - 5\log(h) = -17$ to $-20$, from up to bottom), HOD inside voids is systematically lower for halo masses higher than $\sim 10^{12} \hmsun$ and achieves differences up to $\sim 50\%$.
It is important to note that, for small halo masses, there are no significant differences in the HOD inside voids with respect to the general behaviour.  
This can be interpreted by considering the formation of the first-ranked galaxy of halos nearly independently of the environment, but with the satellite population differing due to their slower formation and accretion in the extreme low-density environment of cosmic voids. This result is in general agreement with \cite{bose_hod_2019} who explore the HOD in TNG300 simulation.

We have further study HOD as a function of number density cuts instead of luminosity thresholds avoiding possible differences arising in the assignment of galaxy luminosities in the two simulated datasets.
The results are given in Sec. \ref{sec:ap2} where it can be seen that the differences between void and global environment remain the same than in Fig. \ref{fig:HOD_voids} using luminosity thresholds.

We have also checked for systematics associated to boundary effects given that HOD is computed in spherical volumes. This was accomplished by randomizing void centre positions and recalculating HOD in these new volumes. For this randomly placed  shperical volumes we recover the HOD behaviour of the overall halo sample, providing confidence in our results.    

\subsection{Dependence on voids radius}
\label{voidradius}
So far, we have seen that there is a clear distinction between the HOD inside cosmic voids as compared to the overall behaviour. 
In this section, we explore a possible dependence on a fundamental parameter of our void sample, namely the void radius. 

\begin{figure*}
\begin{center}
\includegraphics[width=\textwidth]{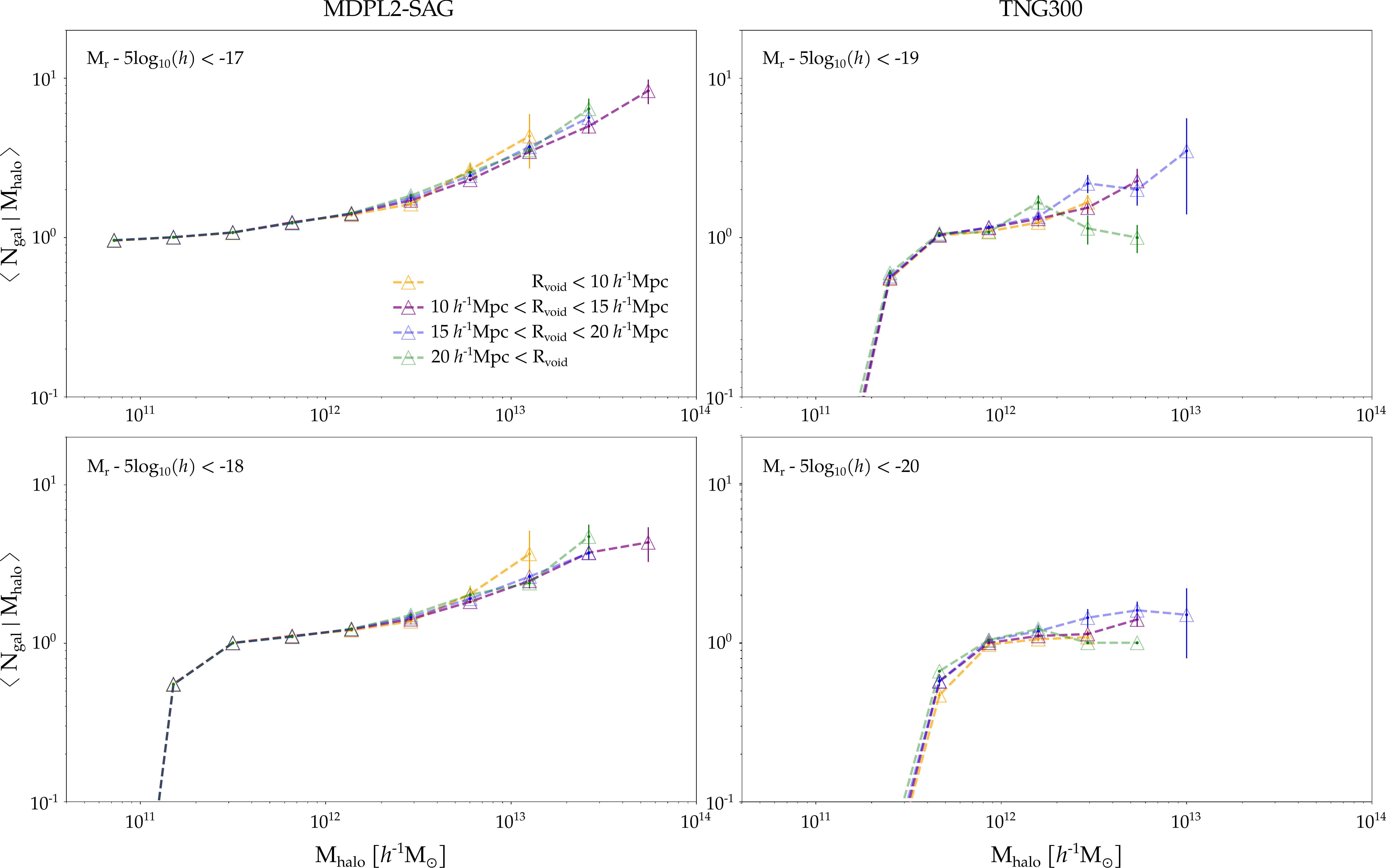}
\end{center}
\caption{\label{HODvsRvoid}HOD in MDPL2-SAG and TNG300 catalogues for different voids radii. 
Left column shows the results for MDPL2-SAG and right column for TNG300 catalogue, for $R_{\rm void} < 10h^{-1}$Mpc (orange), $10h^{-1}$Mpc $<R_{\rm void}< 15h^{-1}$Mpc (purple), $15h^{-1}$Mpc $<R_{\rm void}< 20h^{-1}$Mpc (blue) and $20h^{-1}$Mpc $<R_{\rm void}$ (green). Each panel correspond to a different magnitude limits $M_r-5 \log_{10}(h)$ ranging from -17 to -20. The uncertainties are calculated following the standard jackknife procedure.}
\end{figure*}

We divide our samples into several void radius bins and compute the mean HOD in the different radius bins.  
We define four void sub-samples: $R_{\rm void} < 10h^{-1}$Mpc, $10h^{-1}$Mpc $<R_{\rm void}< 15h^{-1}$Mpc, $15h^{-1}$Mpc $<R_{\rm void}< 20h^{-1}$Mpc and $20h^{-1}$Mpc $<R_{\rm void}$.
In Fig. \ref{HODvsRvoid} we show the results for this analysis. For simplicity, we only present the results for two absolute magnitude thresholds for each galaxy catalogue, MDPL2-SAG on left panels and TNG300 on the right.
The adopted absolute magnitude cuts are labelled on each panel.
We notice that the results are consistent for all studied ranges of absolute magnitudes. 
In the figure, each sub-sample is represented by a different colour as indicated in the key figure (upper left panel).

As it can be clearly seen, within uncertainties there is not a significant dependence of the HOD behaviour on void radius $R_{\rm void}$.
This is somewhat expected because, given our void definition,  up to $R_{\rm void}$ all voids have the same integrated density contrast independently of their size.  

\subsection{Dependence on void large scale environment}

Another way to classify cosmic voids is to use their large-scale environment. %
According to \cite{ceccarelli_clues_2013}, voids can be divided into R-type and S-type voids, where the former are surrounded by large-scale under-dense regions, and the others are embedded into over-dense regions.
This is expressed on void profiles and we use the maximum value of $\Delta$ between $[2R_{\rm void},3R_{\rm void}]$ to classify them.
A void is classified as S-type if the value of $\Delta$ is positive, while if $\Delta$ is negative the void is labelled as R-type. \cite{paz_clues_2013} show that this classification implies that the surrounding region surrounding  R-types void is in expansion and, for S-type voids, in contraction.

\begin{figure}
\begin{center}
\includegraphics[width=\columnwidth]{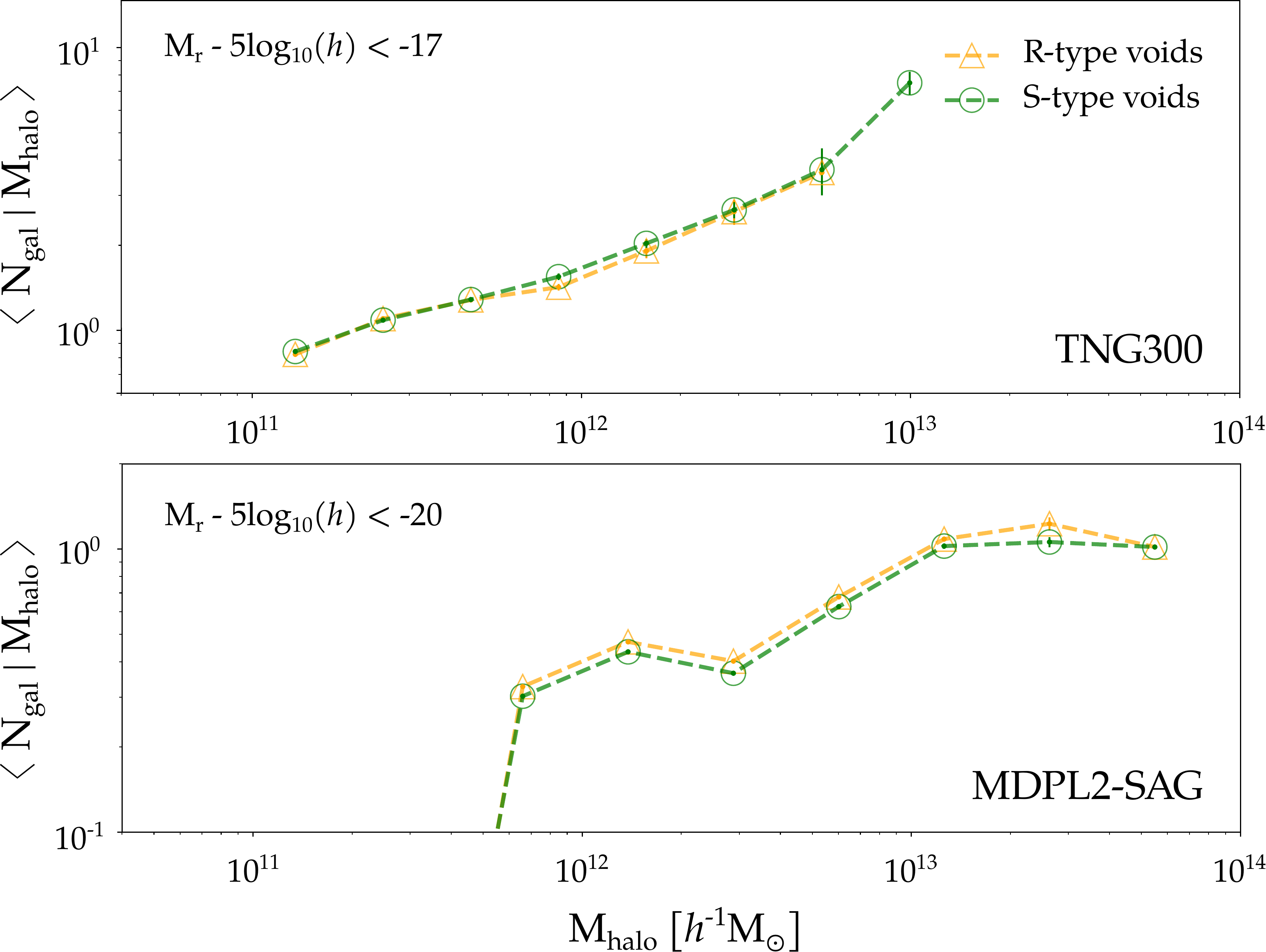}
\end{center}
\caption{\label{HODvsType}HOD in TNG300 and MDPL2-SAG catalogues for different voids classification taking into account their large-scale environment.
The upper panel shows the results for TNG300 and lower panel for the MDPL2-SAG catalogue, for R-type voids (yellow) and S-type voids (green). We show as example only two magnitude limits: $M_r-5\log_{10}(h)=-17$ for TNG300 and $M_r-5\log_{10}(h)=-20$ for MDPL2-SAG. The uncertainties are calculated following the standard jackknife procedure.}
\end{figure}

In this section, we analyse the possibility that these dynamical differences can produce an imprint on the HOD behaviour within the void region.
In Fig. \ref{HODvsType} we show the results of this analysis. For simplicity, only HOD measurements for one absolute magnitude threshold are given for each catalogue, however the results are similar for the whole range of limiting magnitudes studied. 
R-type voids are presented in yellow triangles and S-type in green circles. 
Similarly to \ref{voidradius}, we find no substantial difference between both void sub-samples.  

From these results, we conclude that the effects that drive the changes of the HOD shape inside the voids are present in all cases, independently of void size or surrounding large-scale environment. Here again, we highlight the consistency between the semi-analytic model and the hydro-dynamical simulation results

\section{Stellar mass content} 
\label{sec5}
    
HOD results obtained in the previous sections indicate that, for a dark matter halo with a given mass above $\sim 10^{12} \hmsun$, the number of galaxies decreases if the halo resides inside a cosmic void. 
In this section our aim is to investigate if galaxies in cosmic voids show a different stellar mass content than globally. 
The standard scheme is that the galaxy population can be divided into a central halo galaxy and their satellites. We follow this approach in our analysis.

In order to obtain comparable quantities than those obtained before, we compute the mean stellar mass content for galaxies in haloes of a given mass: $\langle M_\star | M_{\rm halo} \rangle$.
This quantity is calculated for both the central galaxy and the satellite  population considering all galaxies with $M_r - 5\log_{10}(h) < -17$.

\begin{figure*}
\begin{center}
\includegraphics[width=\textwidth]{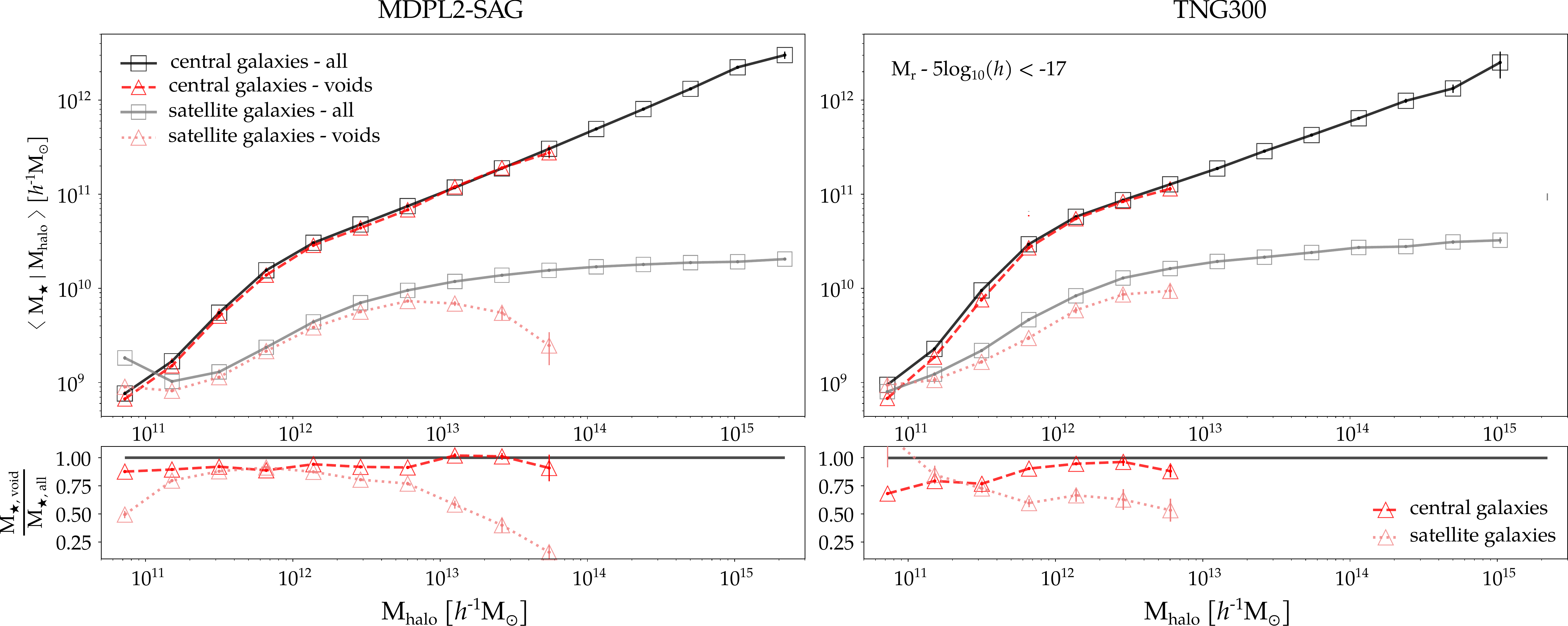}
\end{center}
\caption{\label{fig:mstar}Stellar mass content in central and satellite galaxies for MDPL2-SAG (left) and TNG300 (right) catalogues. Upper panels show $\langle M_\star | M_{\rm halo} \rangle$ as a function of the halo mass for galaxies inside voids (triangles) and all galaxies (squares). 
Bottom panels show the ratio between $M_\star$ for galaxies in voids and the overall population. Dashed lines correspond to central galaxies and dotted lines to the satellite population. Error bars are computed using the standard jackknife procedure.}
\end{figure*}

Fig. \ref{fig:mstar} shows the $\langle M_\star | M_{\rm halo} \rangle$ results for MDPL2-SAG and TNG300 catalogues on the right and left panels, respectively.
As it can be seen in the upper panels, central galaxies in voids (red triangles) have a similar stellar content than the general population of central (dark squares) for both catalogues, meanwhile, satellite galaxies inside voids (light red triangles) show significantly lower values of $M_\star$ with respect to the general behaviour of satellite galaxies (grey squares).
These results can be clearly noticed in the bottom panels, where the ratio $M_{\star,{\rm void}} / M_{\star,{\rm all}}$ is around $\sim 10\%$ for central galaxies (red dashed lines) and $\sim 30 \%$ for satellite galaxies (light red dotted lines).

\begin{figure}
\begin{center}
\includegraphics[width=\columnwidth]{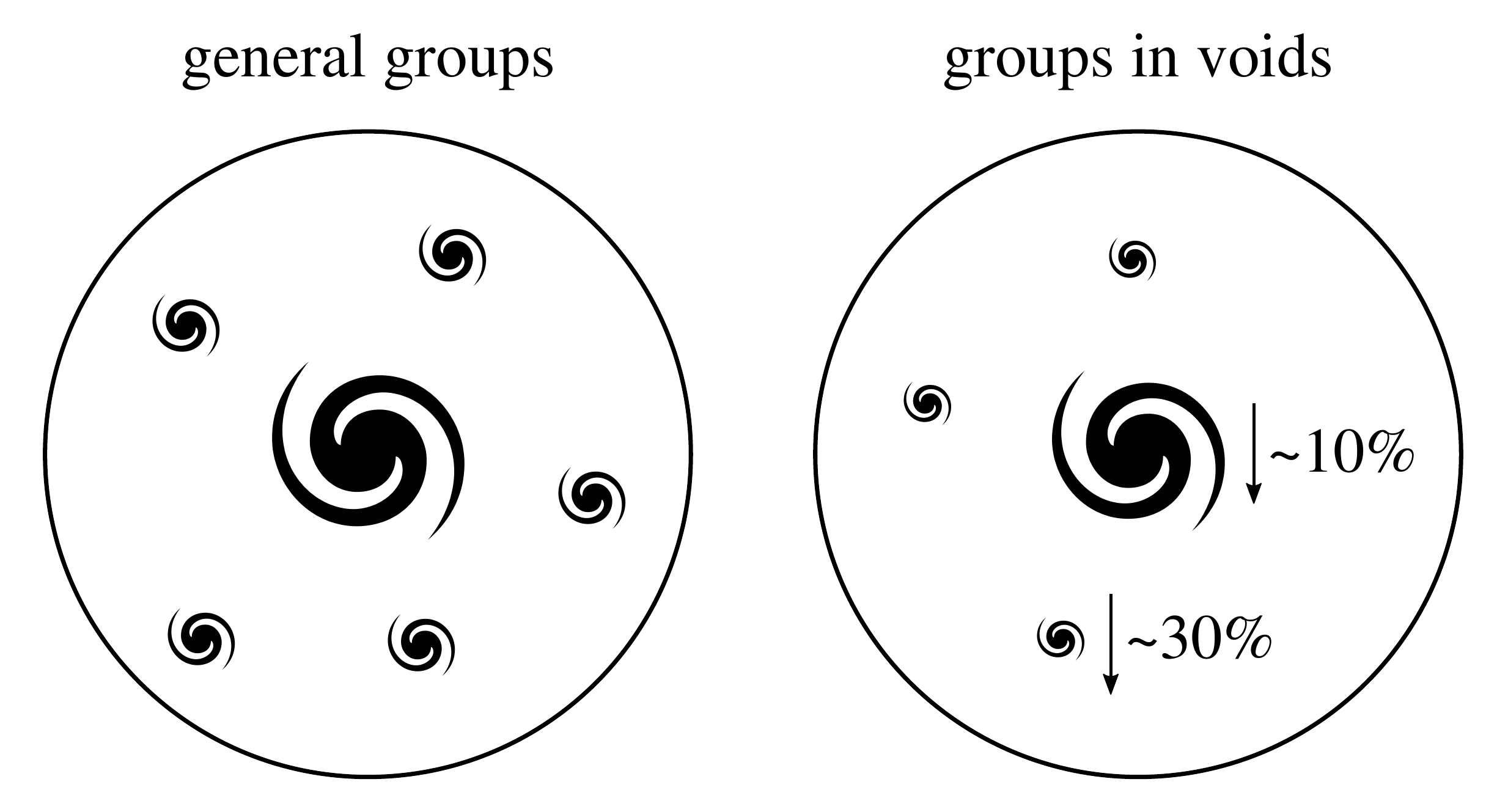}
\end{center}
\caption{\label{fig:esquema}Schematic representation of our main results. Haloes inside cosmic voids have a lower number of member galaxies which also have a lower stellar mass content ($\sim 10 \%$ for central galaxies and $\sim 30\%$ for satellites), for a fixed halo mass and luminosity threshold.}
\end{figure}

It is important to note that HOD and stellar mass content results are consistent for both galaxy catalogues and point out that a halo inside a cosmic void have fewer galaxies, and also each galaxy has a lower stellar mass content. These two main results are schematized in Fig. \ref{fig:esquema}.

\section{Haloes time formation in voids}\label{sec6}

In order to explore possible causes of the differences in the HOD inside cosmic voids, we study in both, MDPL2-SAG and TNG300-1 haloes catalogues, the distribution of $z_{\rm{form}}$, defined as the redshift at which half of the maximum of the halo mass has been accreted onto a halo for the first time. For this aim, we follow the formation history of each halo to determine its maximum mass and the redshift at which it has reached half of this value. We expect that the distribution of this parameter inside voids may differ to that in the complete catalogue and that this different merging history of the haloes populating voids be responsible for its distinct HOD.

\begin{figure}
\begin{center}
\includegraphics[width=\columnwidth]{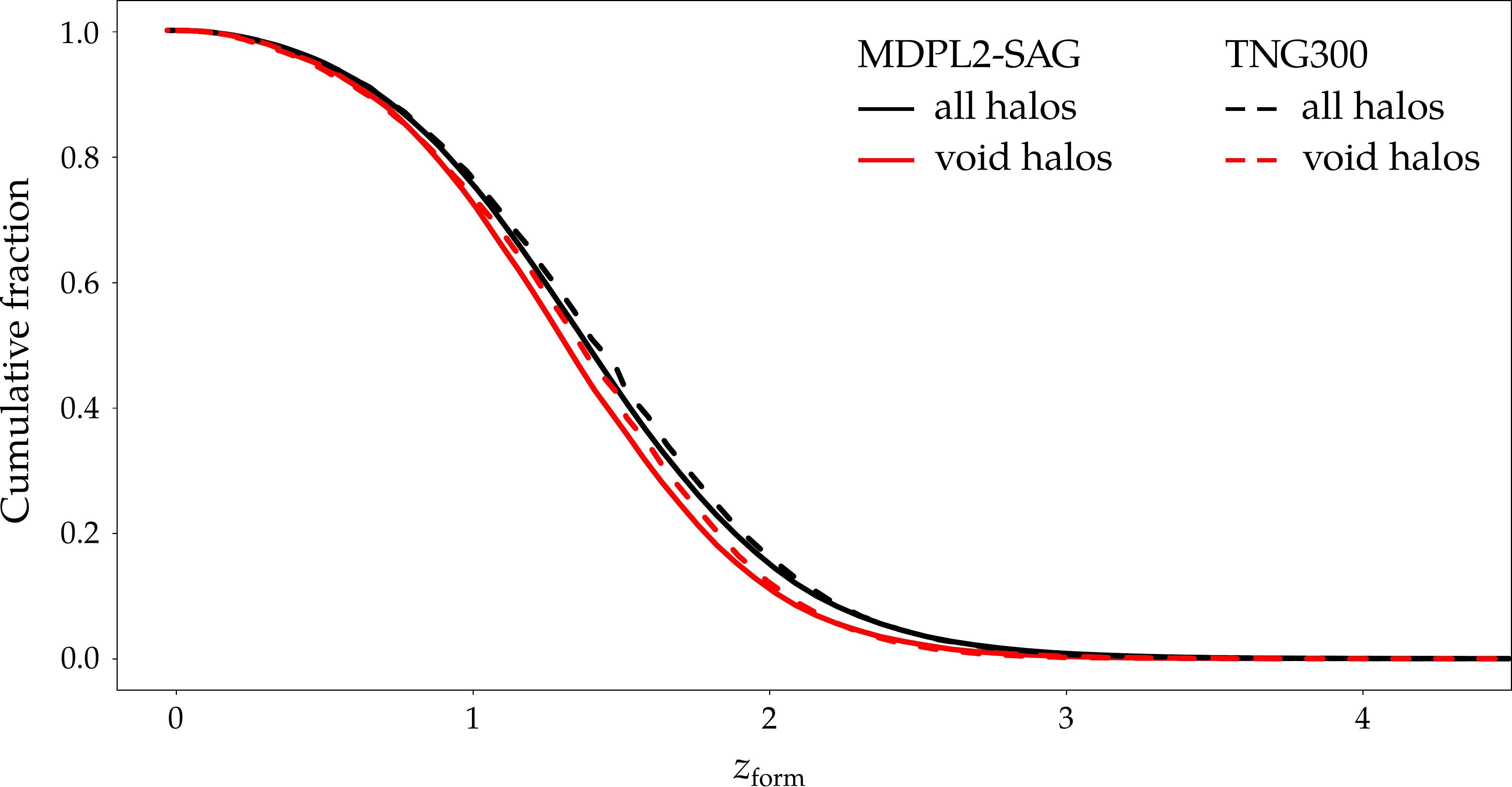}
\end{center}
\caption{\label{fig:halfMass} Cumulative fraction of the formation redshift of haloes, $z_{\rm{form}}$, for MDPL2-SAG (solid lines) and TNG300 (dashed lines) catalogues. Both panels show the distribution for haloes inside voids (red lines) and all haloes (black lines).}
\end{figure}

Fig. \ref{fig:halfMass} shows the cumulative fraction of $z_{\rm{form}}$ for MDPL2-SAG (solid lines) and TNG300 (dashed lines) haloes. In both cases the distributions of $z_{\rm{form}}$ in voids differ from the global behaviour in the same fashion. It is clearer in MDPL2-SAG than in TNG300, showing haloes inside voids (red lines) reaching half of their maximum mass at slightly lower redshifts than elsewhere (black lines). 
For this analysis we select $z_{\rm{form}}$ as the interpolated redshift between the two closest snapshots where haloes achieve 50\% of their maximum mass.

These results are consistent with those obtained in previous sections and provide hints that haloes within cosmic voids have a particular formation history with respect to other environments.


\section{Summary and conclusions}\label{sec7}

The HOD links the galaxies with their host dark matter halo. In this work, we study it behaviour in the extremely low-density environments of cosmic voids. For this purpose, we use two simulated galaxy catalogues derived from the MDPL2-SAG semi-analytic model and the TNG300 hydrodynamic simulation. Although these catalogues have different galaxies luminosity function, void density tracers and are built from very different tools, the results obtained for both catalogues are entirely consistent with each other, so that the conclusions and discussions addressed in this section apply suitably to the two data-sets.

Following our definition of voids, we find a clear dependence of HOD on the environment. The differences increase as the value of the $ \Delta_{\rm{lim}} $ parameter lowers (more empty voids). However, we found that low mass haloes ($\sim 10^{12}\hmsun$) lack variations in HOD, indicating that for these haloes the formation of the central galaxies is nearly independent of the large-scale environment density. For larger masses, haloes in sub-dense regions need more mass than average to increase their number of satellite galaxies.

In this work we use a threshold over-density parameter $\Delta_{\rm{lim}} = -0.9$ to define voids, a usually adopted value that presents the greater differences with respect to the global HOD. Our void measurements show that for different ranges of absolute magnitudes, the HOD is always below the overall results. However, the HOD in voids shows no dependence on void radius nor void large-scale environment. By definition, the interior regions of our voids always present the same integrated mass density, so that the lack of HOD dependence on the other parameters suggests that their primary dependence is on the large-scale density where halo/galaxies reside.

The lower HOD values inside the voids correspond to a smaller number of satellite galaxies per halo. The results of Sec. \ref{sec5} also show that the average stellar mass per satellite is smaller by $\sim 30\%$. Central galaxies are more similar although their stellar mass is smaller by $\sim 10\%$. We conclude that the extreme low-density environment of voids affects more severely the accretion of satellite galaxies generating a population with the fewer satellites and with the lower average stellar mass.

Finally, the results of Sec. \ref{sec6} show the haloes in voids exhibit lower $z_{\rm{form}}$ values, so that void haloes are younger and may have undergone fewer interactions and mergers with other haloes. This fact could explain partially their difficulty in hosting satellite galaxies. 

\begin{acknowledgements}

We kindly thank to the anonymous Referee for his/her very useful comments and suggestions that helped to improve this paper.

This work was partially supported by Agencia Nacional de Promoci\'on Cient\'ifica y Tecnol\'ogica (PICT 2015-3098, PICT 2016-1975), the Consejo Nacional de Investigaciones Científicas y Técnicas (CONICET, Argentina) and the Secretaría de Ciencia y Tecnología de la Universidad Nacional de Córdoba (SeCyT-UNC, Argentina).

The authors gratefully acknowledge the Gauss Centre for Supercomputing e.V. (www.gauss-centre.eu) and the Partnership for Advanced Supercomputing in Europe (PRACE, www.prace-ri.eu) for funding the MultiDark simulation project by providing computing time on the GCS Supercomputer SuperMUC at Leibniz Supercomputing Centre (LRZ, www.lrz.de). The CosmoSim database used in this paper is a service by the Leibniz-Institute for Astrophysics Potsdam (AIP). The MultiDark database was developed in cooperation with the Spanish MultiDark Consolider Project CSD2009-00064.

 The IllustrisTNG project used in this work (TNG300) have been run on the HazelHen Cray XC40-system at the High Performance Computing Center Stuttgart as part of project GCS-ILLU of the Gauss centres for Supercomputing (GCS).

\end{acknowledgements}

\bibliography{references}

\begin{thebibliography}{61}
\expandafter\ifx\csname natexlab\endcsname\relax\def\natexlab#1{#1}\fi

\bibitem[{Artale {et~al.}(2018)Artale, Zehavi, Contreras, \&
  Norberg}]{Artale2018}
Artale, M.~C., Zehavi, I., Contreras, S., \& Norberg, P. 2018, \mnras, 480,
  3978

\bibitem[{{Behroozi} {et~al.}(2013{\natexlab{a}}){Behroozi}, {Wechsler}, \&
  {Wu}}]{behroozi_rockstar_2013}
{Behroozi}, P.~S., {Wechsler}, R.~H., \& {Wu}, H.-Y. 2013{\natexlab{a}}, \apj,
  762, 109

\bibitem[{{Behroozi} {et~al.}(2013{\natexlab{b}}){Behroozi}, {Wechsler}, {Wu},
  {Busha}, {Klypin}, \& {Primack}}]{behroozi_trees_2013}
{Behroozi}, P.~S., {Wechsler}, R.~H., {Wu}, H.-Y., {et~al.} 2013{\natexlab{b}},
  \apj, 763, 18

\bibitem[{Benson {et~al.}(2000)Benson, Cole, Frenk, Baugh, \&
  Lacey}]{Benson2000}
Benson, A., Cole, S., Frenk, C., Baugh, C., \& Lacey, C.~G. 2000, \mnras, 311,
  793

\bibitem[{{Berlind} \& {Weinberg}(2002)}]{Berlind2002}
{Berlind}, A.~A. \& {Weinberg}, D.~H. 2002, \apj, 575, 587

\bibitem[{Berlind {et~al.}(2003)Berlind, Weinberg, Benson, Baugh, Cole,
  Dav{\'e}, Frenk, Jenkins, Katz, \& Lacey}]{Berlind2003}
Berlind, A.~A., Weinberg, D.~H., Benson, A.~J., {et~al.} 2003, \apj, 593, 1

\bibitem[{{Bose} {et~al.}(2019){Bose}, {Eisenstein}, {Hernquist}, {Pillepich},
  {Nelson}, {Marinacci}, {Springel}, \& {Vogelsberger}}]{bose_hod_2019}
{Bose}, S., {Eisenstein}, D.~J., {Hernquist}, L., {et~al.} 2019, \mnras, 490,
  5693

\bibitem[{{Cautun} {et~al.}(2018){Cautun}, {Paillas}, {Cai}, {Bose}, {Armijo},
  {Li}, \& {Padilla}}]{Cautun2018}
{Cautun}, M., {Paillas}, E., {Cai}, Y.-C., {et~al.} 2018, \mnras, 476, 3195

\bibitem[{{Ceccarelli} {et~al.}(2008){Ceccarelli}, {Padilla}, \&
  {Lambas}}]{Ceccarelli2008}
{Ceccarelli}, L., {Padilla}, N., \& {Lambas}, D.~G. 2008, \mnras, 390, L9

\bibitem[{{Ceccarelli} {et~al.}(2006){Ceccarelli}, {Padilla}, {Valotto}, \&
  {Lambas}}]{ceccarelli_voids_2006}
{Ceccarelli}, L., {Padilla}, N.~D., {Valotto}, C., \& {Lambas}, D.~G. 2006,
  \mnras, 373, 1440

\bibitem[{{Ceccarelli} {et~al.}(2013){Ceccarelli}, {Paz}, {Lares}, {Padilla},
  \& {Lambas}}]{ceccarelli_clues_2013}
{Ceccarelli}, L., {Paz}, D., {Lares}, M., {Padilla}, N., \& {Lambas}, D.~G.
  2013, \mnras, 434, 1435

\bibitem[{{Colberg} {et~al.}(2008){Colberg}, {Pearce}, {Foster}, {Platen},
  {Brunino}, {Neyrinck}, {Basilakos}, {Fairall}, {Feldman}, {Gottl{\"o}ber},
  {Hahn}, {Hoyle}, {M{\"u}ller}, {Nelson}, {Plionis}, {Porciani}, {Shandarin},
  {Vogeley}, \& {van de Weygaert}}]{Colberg2008}
{Colberg}, J.~M., {Pearce}, F., {Foster}, C., {et~al.} 2008, \mnras, 387, 933

\bibitem[{Cooray \& Sheth(2002)}]{Cooray2002}
Cooray, A. \& Sheth, R. 2002, Physics reports, 372, 1

\bibitem[{Cora(2006)}]{cora_sag_2006}
Cora, S.~A. 2006, \mnras, 368, 1540

\bibitem[{{Cora} {et~al.}(2018){Cora}, {Vega-Mart{\'\i}nez}, {Hough}, {Ruiz},
  {Orsi}, {Mu{\~n}oz Arancibia}, {Gargiulo}, {Collacchioni}, {Padilla},
  {Gottl{\"o}ber}, \& {Yepes}}]{cora_sag_2018}
{Cora}, S.~A., {Vega-Mart{\'\i}nez}, C.~A., {Hough}, T., {et~al.} 2018, \mnras,
  479, 2

\bibitem[{{Gargiulo} {et~al.}(2015){Gargiulo}, {Cora}, {Padilla}, {Mu{\~n}oz
  Arancibia}, {Ruiz}, {Orsi}, {Tecce}, {Weidner}, \&
  {Bruzual}}]{gargiulo_sag_2015}
{Gargiulo}, I.~D., {Cora}, S.~A., {Padilla}, N.~D., {et~al.} 2015, \mnras, 446,
  3820

\bibitem[{{Guo} {et~al.}(2015){Guo}, {Zheng}, {Zehavi}, {Behroozi}, {Chuang},
  {Comparat}, {Favole}, {Gottloeber}, {Klypin}, {Prada}, {Weinberg}, \&
  {Yepes}}]{Guo2015}
{Guo}, H., {Zheng}, Z., {Zehavi}, I., {et~al.} 2015, \mnras, 453, 4368

\bibitem[{{Hoyle} {et~al.}(2005){Hoyle}, {Rojas}, {Vogeley}, \&
  {Brinkmann}}]{Hoyle2005}
{Hoyle}, F., {Rojas}, R.~R., {Vogeley}, M.~S., \& {Brinkmann}, J. 2005, \apj,
  620, 618

\bibitem[{{Hoyle} {et~al.}(2012){Hoyle}, {Vogeley}, \& {Pan}}]{Hoyle2012}
{Hoyle}, F., {Vogeley}, M.~S., \& {Pan}, D. 2012, \mnras, 426, 3041

\bibitem[{Jing {et~al.}(1998)Jing, Mo, \& B{\"o}rner}]{Jing1998}
Jing, Y., Mo, H., \& B{\"o}rner, G. 1998, \apj, 494, 1

\bibitem[{{Klypin} {et~al.}(2016){Klypin}, {Yepes}, {Gottl{\"o}ber}, {Prada},
  \& {He{\ss}}}]{klypin_multidark_2016}
{Klypin}, A., {Yepes}, G., {Gottl{\"o}ber}, S., {Prada}, F., \& {He{\ss}}, S.
  2016, \mnras, 457, 4340

\bibitem[{{Knebe} {et~al.}(2018){Knebe}, {Stoppacher}, {Prada}, {Behrens},
  {Benson}, {Cora}, {Croton}, {Padilla}, {Ruiz}, {Sinha}, {Stevens},
  {Vega-Mart{\'\i}nez}, {Behroozi}, {Gonzalez-Perez}, {Gottl{\"o}ber},
  {Klypin}, {Yepes}, {Enke}, {Libeskind}, {Riebe}, \&
  {Steinmetz}}]{knebe_multidark_2018}
{Knebe}, A., {Stoppacher}, D., {Prada}, F., {et~al.} 2018, \mnras, 474, 5206

\bibitem[{{Kravtsov} {et~al.}(2004){Kravtsov}, {Berlind}, {Wechsler}, {Klypin},
  {Gottl{\"o}ber}, {Allgood}, \& {Primack}}]{Kravtsov2004}
{Kravtsov}, A.~V., {Berlind}, A.~A., {Wechsler}, R.~H., {et~al.} 2004, \apj,
  609, 35

\bibitem[{Lagos {et~al.}(2008)Lagos, Cora, \& Padilla}]{lagos_sag_2008}
Lagos, C. d.~P., Cora, S.~A., \& Padilla, N.~D. 2008, \mnras, 388, 587

\bibitem[{Ma \& Fry(2000)}]{Ma2000}
Ma, C.-P. \& Fry, J.~N. 2000, \apj, 543, 503

\bibitem[{{Marinacci} {et~al.}(2018){Marinacci}, {Vogelsberger}, {Pakmor},
  {Torrey}, {Springel}, {Hernquist}, {Nelson}, {Weinberger}, {Pillepich},
  {Naiman}, \& {Genel}}]{marinacci_tng_2018}
{Marinacci}, F., {Vogelsberger}, M., {Pakmor}, R., {et~al.} 2018, \mnras, 480,
  5113

\bibitem[{{Naiman} {et~al.}(2018){Naiman}, {Pillepich}, {Springel},
  {Ramirez-Ruiz}, {Torrey}, {Vogelsberger}, {Pakmor}, {Nelson}, {Marinacci},
  {Hernquist}, {Weinberger}, \& {Genel}}]{naiman_tng_2018}
{Naiman}, J.~P., {Pillepich}, A., {Springel}, V., {et~al.} 2018, \mnras, 477,
  1206

\bibitem[{{Nelson} {et~al.}(2018){Nelson}, {Pillepich}, {Springel},
  {Weinberger}, {Hernquist}, {Pakmor}, {Genel}, {Torrey}, {Vogelsberger},
  {Kauffmann}, {Marinacci}, \& {Naiman}}]{nelson_tng_2018}
{Nelson}, D., {Pillepich}, A., {Springel}, V., {et~al.} 2018, \mnras, 475, 624

\bibitem[{{Padilla} {et~al.}(2005){Padilla}, {Ceccarelli}, \&
  {Lambas}}]{padilla_void_2005}
{Padilla}, N.~D., {Ceccarelli}, L., \& {Lambas}, D.~G. 2005, \mnras, 363, 977

\bibitem[{{Pan} {et~al.}(2012){Pan}, {Vogeley}, {Hoyle}, {Choi}, \&
  {Park}}]{Pan2012}
{Pan}, D.~C., {Vogeley}, M.~S., {Hoyle}, F., {Choi}, Y.-Y., \& {Park}, C. 2012,
  \mnras, 421, 926

\bibitem[{{Patiri} {et~al.}(2006){Patiri}, {Prada}, {Holtzman}, {Klypin}, \&
  {Betancort-Rijo}}]{Patiri2006}
{Patiri}, S.~G., {Prada}, F., {Holtzman}, J., {Klypin}, A., \&
  {Betancort-Rijo}, J. 2006, \mnras, 372, 1710

\bibitem[{{Paz} {et~al.}(2013){Paz}, {Lares}, {Ceccarelli}, {Padilla}, \&
  {Lambas}}]{paz_clues_2013}
{Paz}, D., {Lares}, M., {Ceccarelli}, L., {Padilla}, N., \& {Lambas}, D.~G.
  2013, \mnras, 436, 3480

\bibitem[{Peacock \& Smith(2000)}]{Peacock2000}
Peacock, J. \& Smith, R. 2000, \mnras, 318, 1144

\bibitem[{{Peebles}(1980)}]{Peebles1980}
{Peebles}, P.~J.~E. 1980, {The large-scale structure of the universe}

\bibitem[{{Pillepich} {et~al.}(2018{\natexlab{a}}){Pillepich}, {Nelson},
  {Hernquist}, {Springel}, {Pakmor}, {Torrey}, {Weinberger}, {Genel}, {Naiman},
  {Marinacci}, \& {Vogelsberger}}]{pillepich_tng_2018}
{Pillepich}, A., {Nelson}, D., {Hernquist}, L., {et~al.} 2018{\natexlab{a}},
  \mnras, 475, 648

\bibitem[{{Pillepich} {et~al.}(2018{\natexlab{b}}){Pillepich}, {Springel},
  {Nelson}, {Genel}, {Naiman}, {Pakmor}, {Hernquist}, {Torrey}, {Vogelsberger},
  {Weinberger}, \& {Marinacci}}]{pillepich_tng0_2018}
{Pillepich}, A., {Springel}, V., {Nelson}, D., {et~al.} 2018{\natexlab{b}},
  \mnras, 473, 4077

\bibitem[{{Planck Collaboration} {et~al.}(2014){Planck Collaboration}, {Ade},
  {Aghanim}, {Armitage-Caplan}, {Arnaud}, {Ashdown}, {Atrio-Barand ela},
  {Aumont}, {Baccigalupi}, {Banday}, {Barreiro}, {Bartlett}, {Battaner},
  {Benabed}, {Beno{\^\i}t}, {Benoit-L{\'e}vy}, {Bernard}, {Bersanelli},
  {Bielewicz}, {Bobin}, {Bock}, {Bonaldi}, {Bond}, {Borrill}, {Bouchet},
  {Bridges}, {Bucher}, {Burigana}, {Butler}, {Calabrese}, {Cappellini},
  {Cardoso}, {Catalano}, {Challinor}, {Chamballu}, {Chary}, {Chen}, {Chiang},
  {Chiang}, {Christensen}, {Church}, {Clements}, {Colombi}, {Colombo},
  {Couchot}, {Coulais}, {Crill}, {Curto}, {Cuttaia}, {Danese}, {Davies},
  {Davis}, {de Bernardis}, {de Rosa}, {de Zotti}, {Delabrouille}, {Delouis},
  {D{\'e}sert}, {Dickinson}, {Diego}, {Dolag}, {Dole}, {Donzelli}, {Dor{\'e}},
  {Douspis}, {Dunkley}, {Dupac}, {Efstathiou}, {Elsner}, {En{\ss}lin},
  {Eriksen}, {Finelli}, {Forni}, {Frailis}, {Fraisse}, {Franceschi}, {Gaier},
  {Galeotta}, {Galli}, {Ganga}, {Giard}, {Giardino}, {Giraud-H{\'e}raud},
  {Gjerl{\o}w}, {Gonz{\'a}lez-Nuevo}, {G{\'o}rski}, {Gratton}, {Gregorio},
  {Gruppuso}, {Gudmundsson}, {Haissinski}, {Hamann}, {Hansen}, {Hanson},
  {Harrison}, {Henrot-Versill{\'e}}, {Hern{\'a}ndez-Monteagudo}, {Herranz},
  {Hildebrand t}, {Hivon}, {Hobson}, {Holmes}, {Hornstrup}, {Hou}, {Hovest},
  {Huffenberger}, {Jaffe}, {Jaffe}, {Jewell}, {Jones}, {Juvela},
  {Keih{\"a}nen}, {Keskitalo}, {Kisner}, {Kneissl}, {Knoche}, {Knox}, {Kunz},
  {Kurki-Suonio}, {Lagache}, {L{\"a}hteenm{\"a}ki}, {Lamarre}, {Lasenby},
  {Lattanzi}, {Laureijs}, {Lawrence}, {Leach}, {Leahy}, {Leonardi},
  {Le{\'o}n-Tavares}, {Lesgourgues}, {Lewis}, {Liguori}, {Lilje},
  {Linden-V{\o}rnle}, {L{\'o}pez-Caniego}, {Lubin}, {Mac{\'\i}as-P{\'e}rez},
  {Maffei}, {Maino}, {Mand olesi}, {Maris}, {Marshall}, {Martin},
  {Mart{\'\i}nez-Gonz{\'a}lez}, {Masi}, {Massardi}, {Matarrese}, {Matthai},
  {Mazzotta}, {Meinhold}, {Melchiorri}, {Melin}, {Mendes}, {Menegoni},
  {Mennella}, {Migliaccio}, {Millea}, {Mitra}, {Miville-Desch{\^e}nes},
  {Moneti}, {Montier}, {Morgante}, {Mortlock}, {Moss}, {Munshi}, {Murphy},
  {Naselsky}, {Nati}, {Natoli}, {Netterfield}, {N{\o}rgaard-Nielsen},
  {Noviello}, {Novikov}, {Novikov}, {O'Dwyer}, {Osborne}, {Oxborrow}, {Paci},
  {Pagano}, {Pajot}, {Paladini}, {Paoletti}, {Partridge}, {Pasian},
  {Patanchon}, {Pearson}, {Pearson}, {Peiris}, {Perdereau}, {Perotto},
  {Perrotta}, {Pettorino}, {Piacentini}, {Piat}, {Pierpaoli}, {Pietrobon},
  {Plaszczynski}, {Platania}, {Pointecouteau}, {Polenta}, {Ponthieu}, {Popa},
  {Poutanen}, {Pratt}, {Pr{\'e}zeau}, {Prunet}, {Puget}, {Rachen}, {Reach},
  {Rebolo}, {Reinecke}, {Remazeilles}, {Renault}, {Ricciardi}, {Riller},
  {Ristorcelli}, {Rocha}, {Rosset}, {Roudier}, {Rowan-Robinson},
  {Rubi{\~n}o-Mart{\'\i}n}, {Rusholme}, {Sandri}, {Santos}, {Savelainen},
  {Savini}, {Scott}, {Seiffert}, {Shellard}, {Spencer}, {Starck}, {Stolyarov},
  {Stompor}, {Sudiwala}, {Sunyaev}, {Sureau}, {Sutton}, {Suur-Uski}, {Sygnet},
  {Tauber}, {Tavagnacco}, {Terenzi}, {Toffolatti}, {Tomasi}, {Tristram},
  {Tucci}, {Tuovinen}, {T{\"u}rler}, {Umana}, {Valenziano}, {Valiviita}, {Van
  Tent}, {Vielva}, {Villa}, {Vittorio}, {Wade}, {Wandelt}, {Wehus}, {White},
  {White}, {Wilkinson}, {Yvon}, {Zacchei}, \& {Zonca}}]{planck_2014}
{Planck Collaboration}, {Ade}, P.~A.~R., {Aghanim}, N., {et~al.} 2014, \aap,
  571, A16

\bibitem[{{Planck Collaboration} {et~al.}(2016){Planck Collaboration}, {Ade},
  {Aghanim}, {Arnaud}, {Ashdown}, {Aumont}, {Baccigalupi}, {Banday},
  {Barreiro}, {Bartlett}, {Bartolo}, {Battaner}, {Battye}, {Benabed},
  {Beno{\^\i}t}, {Benoit-L{\'e}vy}, {Bernard}, {Bersanelli}, {Bielewicz},
  {Bock}, {Bonaldi}, {Bonavera}, {Bond}, {Borrill}, {Bouchet}, {Boulanger},
  {Bucher}, {Burigana}, {Butler}, {Calabrese}, {Cardoso}, {Catalano},
  {Challinor}, {Chamballu}, {Chary}, {Chiang}, {Chluba}, {Christensen},
  {Church}, {Clements}, {Colombi}, {Colombo}, {Combet}, {Coulais}, {Crill},
  {Curto}, {Cuttaia}, {Danese}, {Davies}, {Davis}, {de Bernardis}, {de Rosa},
  {de Zotti}, {Delabrouille}, {D{\'e}sert}, {Di Valentino}, {Dickinson},
  {Diego}, {Dolag}, {Dole}, {Donzelli}, {Dor{\'e}}, {Douspis}, {Ducout},
  {Dunkley}, {Dupac}, {Efstathiou}, {Elsner}, {En{\ss}lin}, {Eriksen},
  {Farhang}, {Fergusson}, {Finelli}, {Forni}, {Frailis}, {Fraisse},
  {Franceschi}, {Frejsel}, {Galeotta}, {Galli}, {Ganga}, {Gauthier}, {Gerbino},
  {Ghosh}, {Giard}, {Giraud-H{\'e}raud}, {Giusarma}, {Gjerl{\o}w},
  {Gonz{\'a}lez-Nuevo}, {G{\'o}rski}, {Gratton}, {Gregorio}, {Gruppuso},
  {Gudmundsson}, {Hamann}, {Hansen}, {Hanson}, {Harrison}, {Helou},
  {Henrot-Versill{\'e}}, {Hern{\'a}ndez-Monteagudo}, {Herranz}, {Hildebrand t},
  {Hivon}, {Hobson}, {Holmes}, {Hornstrup}, {Hovest}, {Huang}, {Huffenberger},
  {Hurier}, {Jaffe}, {Jaffe}, {Jones}, {Juvela}, {Keih{\"a}nen}, {Keskitalo},
  {Kisner}, {Kneissl}, {Knoche}, {Knox}, {Kunz}, {Kurki-Suonio}, {Lagache},
  {L{\"a}hteenm{\"a}ki}, {Lamarre}, {Lasenby}, {Lattanzi}, {Lawrence}, {Leahy},
  {Leonardi}, {Lesgourgues}, {Levrier}, {Lewis}, {Liguori}, {Lilje},
  {Linden-V{\o}rnle}, {L{\'o}pez-Caniego}, {Lubin}, {Mac{\'\i}as-P{\'e}rez},
  {Maggio}, {Maino}, {Mandolesi}, {Mangilli}, {Marchini}, {Maris}, {Martin},
  {Martinelli}, {Mart{\'\i}nez-Gonz{\'a}lez}, {Masi}, {Matarrese}, {McGehee},
  {Meinhold}, {Melchiorri}, {Melin}, {Mendes}, {Mennella}, {Migliaccio},
  {Millea}, {Mitra}, {Miville-Desch{\^e}nes}, {Moneti}, {Montier}, {Morgante},
  {Mortlock}, {Moss}, {Munshi}, {Murphy}, {Naselsky}, {Nati}, {Natoli},
  {Netterfield}, {N{\o}rgaard-Nielsen}, {Noviello}, {Novikov}, {Novikov},
  {Oxborrow}, {Paci}, {Pagano}, {Pajot}, {Paladini}, {Paoletti}, {Partridge},
  {Pasian}, {Patanchon}, {Pearson}, {Perdereau}, {Perotto}, {Perrotta},
  {Pettorino}, {Piacentini}, {Piat}, {Pierpaoli}, {Pietrobon}, {Plaszczynski},
  {Pointecouteau}, {Polenta}, {Popa}, {Pratt}, {Pr{\'e}zeau}, {Prunet},
  {Puget}, {Rachen}, {Reach}, {Rebolo}, {Reinecke}, {Remazeilles}, {Renault},
  {Renzi}, {Ristorcelli}, {Rocha}, {Rosset}, {Rossetti}, {Roudier},
  {Rouill{\'e} d'Orfeuil}, {Rowan-Robinson}, {Rubi{\~n}o-Mart{\'\i}n},
  {Rusholme}, {Said}, {Salvatelli}, {Salvati}, {Sandri}, {Santos},
  {Savelainen}, {Savini}, {Scott}, {Seiffert}, {Serra}, {Shellard}, {Spencer},
  {Spinelli}, {Stolyarov}, {Stompor}, {Sudiwala}, {Sunyaev}, {Sutton},
  {Suur-Uski}, {Sygnet}, {Tauber}, {Terenzi}, {Toffolatti}, {Tomasi},
  {Tristram}, {Trombetti}, {Tucci}, {Tuovinen}, {T{\"u}rler}, {Umana},
  {Valenziano}, {Valiviita}, {Van Tent}, {Vielva}, {Villa}, {Wade}, {Wandelt},
  {Wehus}, {White}, {White}, {Wilkinson}, {Yvon}, {Zacchei}, \&
  {Zonca}}]{planck_2016}
{Planck Collaboration}, {Ade}, P.~A.~R., {Aghanim}, N., {et~al.} 2016, \aap,
  594, A13

\bibitem[{Pujol \& Gazta{\~n}aga(2014)}]{Pujol2014}
Pujol, A. \& Gazta{\~n}aga, E. 2014, \mnras, 442, 1930

\bibitem[{Pujol {et~al.}(2017)Pujol, Hoffmann, Jim{\'e}nez, \&
  Gazta{\~n}aga}]{Pujol2017}
Pujol, A., Hoffmann, K., Jim{\'e}nez, N., \& Gazta{\~n}aga, E. 2017, \aap, 598,
  A103

\bibitem[{{Riebe} {et~al.}(2013){Riebe}, {Partl}, {Enke}, {Forero-Romero},
  {Gottl{\"o}ber}, {Klypin}, {Lemson}, {Prada}, {Primack}, {Steinmetz}, \&
  {Turchaninov}}]{riebe_multidark_2013}
{Riebe}, K., {Partl}, A.~M., {Enke}, H., {et~al.} 2013, Astronomische
  Nachrichten, 334, 691

\bibitem[{{Rodriguez} {et~al.}(2015){Rodriguez}, {Merch{\'a}n}, \&
  {Sgr{\'o}}}]{Rodriguez2015}
{Rodriguez}, F., {Merch{\'a}n}, M., \& {Sgr{\'o}}, M.~A. 2015, \aap, 580, A86

\bibitem[{{Rojas} {et~al.}(2004){Rojas}, {Vogeley}, {Hoyle}, \&
  {Brinkmann}}]{Rojas2004}
{Rojas}, R.~R., {Vogeley}, M.~S., {Hoyle}, F., \& {Brinkmann}, J. 2004, \apj,
  617, 50

\bibitem[{{Rojas} {et~al.}(2005){Rojas}, {Vogeley}, {Hoyle}, \&
  {Brinkmann}}]{Rojas2005}
{Rojas}, R.~R., {Vogeley}, M.~S., {Hoyle}, F., \& {Brinkmann}, J. 2005, \apj,
  624, 571

\bibitem[{{Ruiz} {et~al.}(2019){Ruiz}, {Alfaro}, \& {Garcia
  Lambas}}]{ruiz_into_2019}
{Ruiz}, A.~N., {Alfaro}, I.~G., \& {Garcia Lambas}, D. 2019, \mnras, 483, 4070

\bibitem[{{Ruiz} {et~al.}(2015{\natexlab{a}}){Ruiz}, {Cora}, {Padilla},
  {Dom{\'\i}nguez}, {Vega-Mart{\'\i}nez}, {Tecce}, {Orsi}, {Yaryura},
  {Garc{\'\i}a Lambas}, {Gargiulo}, \& {Mu{\~n}oz Arancibia}}]{ruiz_sag_2015}
{Ruiz}, A.~N., {Cora}, S.~A., {Padilla}, N.~D., {et~al.} 2015{\natexlab{a}},
  \apj, 801, 139

\bibitem[{{Ruiz} {et~al.}(2015{\natexlab{b}}){Ruiz}, {Paz}, {Lares},
  {Luparello}, {Ceccarelli}, \& {Lambas}}]{ruiz_void_2015}
{Ruiz}, A.~N., {Paz}, D.~J., {Lares}, M., {et~al.} 2015{\natexlab{b}}, \mnras,
  448, 1471

\bibitem[{Scoccimarro {et~al.}(2001)Scoccimarro, Sheth, Hui, \&
  Jain}]{Scoccimarro2001}
Scoccimarro, R., Sheth, R.~K., Hui, L., \& Jain, B. 2001, \apj, 546, 20

\bibitem[{Seljak(2000)}]{Seljak2000}
Seljak, U. 2000, \mnras, 318, 203

\bibitem[{{Springel}(2010)}]{springel_arepo_2010}
{Springel}, V. 2010, \mnras, 401, 791

\bibitem[{{Springel} {et~al.}(2018){Springel}, {Pakmor}, {Pillepich},
  {Weinberger}, {Nelson}, {Hernquist}, {Vogelsberger}, {Genel}, {Torrey},
  {Marinacci}, \& {Naiman}}]{springel_tng_2018}
{Springel}, V., {Pakmor}, R., {Pillepich}, A., {et~al.} 2018, \mnras, 475, 676

\bibitem[{Tecce {et~al.}(2010)Tecce, Cora, Tissera, Abadi, \&
  Lagos}]{tecce_sag_2010}
Tecce, T.~E., Cora, S.~A., Tissera, P.~B., Abadi, M.~G., \& Lagos, C. d.~P.
  2010, \mnras, 408, 2008

\bibitem[{Van Den~Bosch {et~al.}(2003)Van Den~Bosch, Yang, \&
  Mo}]{Vandenbosch2003}
Van Den~Bosch, F.~C., Yang, X., \& Mo, H. 2003, \mnras, 340, 771

\bibitem[{{Weinberger} {et~al.}(2017){Weinberger}, {Springel}, {Hernquist},
  {Pillepich}, {Marinacci}, {Pakmor}, {Nelson}, {Genel}, {Vogelsberger},
  {Naiman}, \& {Torrey}}]{weinberger_tng_2017}
{Weinberger}, R., {Springel}, V., {Hernquist}, L., {et~al.} 2017, \mnras, 465,
  3291

\bibitem[{White \& Rees(1978)}]{White1978}
White, S.~D. \& Rees, M.~J. 1978, \mnras, 183, 341

\bibitem[{{Yang} {et~al.}(2007){Yang}, {Mo}, {van den Bosch}, {Pasquali}, {Li},
  \& {Barden}}]{Yang2007}
{Yang}, X., {Mo}, H.~J., {van den Bosch}, F.~C., {et~al.} 2007, \apj, 671, 153

\bibitem[{Zehavi {et~al.}(2018)Zehavi, Contreras, Padilla, Smith, Baugh, \&
  Norberg}]{Zehavi2018}
Zehavi, I., Contreras, S., Padilla, N., {et~al.} 2018, \apj, 853, 84

\bibitem[{Zehavi {et~al.}(2011)Zehavi, Zheng, Weinberg, Blanton, Bahcall,
  Berlind, Brinkmann, Frieman, Gunn, Lupton, {et~al.}}]{Zehavi2011}
Zehavi, I., Zheng, Z., Weinberg, D.~H., {et~al.} 2011, \apj, 736, 59

\bibitem[{Zentner {et~al.}(2005)Zentner, Berlind, Bullock, Kravtsov, \&
  Wechsler}]{Zentner2005}
Zentner, A.~R., Berlind, A.~A., Bullock, J.~S., Kravtsov, A.~V., \& Wechsler,
  R.~H. 2005, \apj, 624, 505

\bibitem[{{Zheng} {et~al.}(2005){Zheng}, {Berlind}, {Weinberg}, {Benson},
  {Baugh}, {Cole}, {Dav{\'e}}, {Frenk}, {Katz}, \& {Lacey}}]{Zheng2005}
{Zheng}, Z., {Berlind}, A.~A., {Weinberg}, D.~H., {et~al.} 2005, \apj, 633, 791

\bibitem[{Zheng \& Weinberg(2007)}]{Zheng2007}
Zheng, Z. \& Weinberg, D.~H. 2007, \apj, 659, 1

\end{thebibliography}


\begin{appendix}

\section{Impact of number density in HOD deterinations}

As described in Sec. \ref{2.2}, by imposing the same absolute magnitude cut in both galaxy catalogues, and given the differences in the luminosity functions, we obtain galaxy samples with different number densities. As a consequence, HOD are measured in galaxy populations 
with different number densities. Besides, voids identified using these galaxy tracers have also different number densities as a function of void sizes. 
In this appendix we explore the consequences in HOD measurements when we impose void and galaxy catalogues to have the same number density in both simulations. 

\subsection{Equal number density of voids limiting by void radius}
\label{sec:ap1}

\begin{figure}[h]
\begin{center}
\includegraphics[width=\columnwidth]{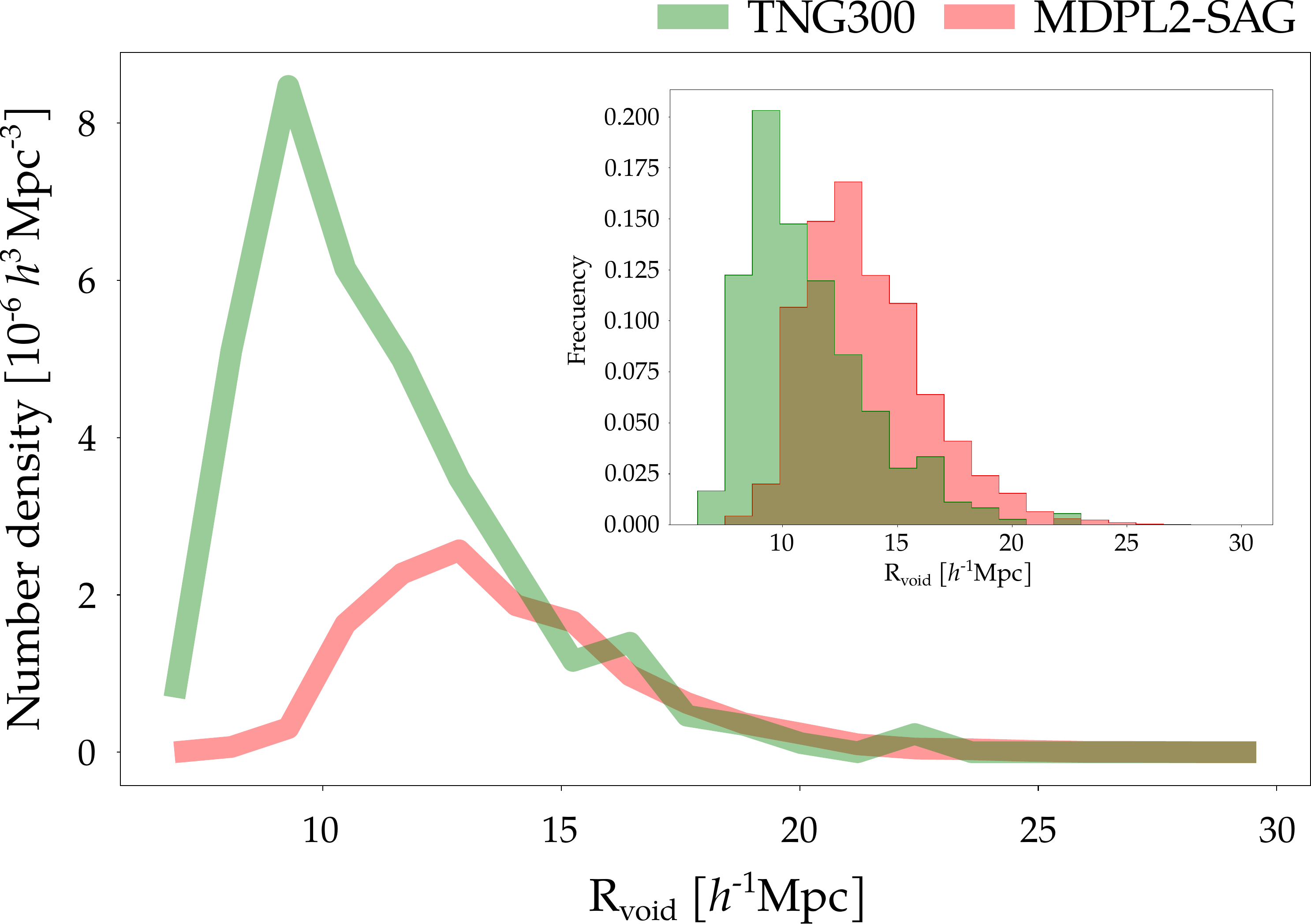}
\end{center}
\caption{\label{fig:VoidDensity} Volume number density of voids as a function of void sizes. Green and red curves correspond to TNG300 and MDPL2-SAG catalogues, respectively. The inset figure show the size distributions for both catalogues.}
\end{figure}

Fig. \ref{fig:VoidDensity} shows the volume number density of voids as a function of radius for TNG300 (green) and MDPL2-SAG (red). In the inset figure we also present the radii distribution for both voids catalogues. 
As it can be seen, the number densities of voids in both catalogues are different, however it is clear from the figure that if we take voids with $R_{\rm{void}} \sim 13 h^{-1}$Mpc, both samples are not only similar in density but also complete in void sizes.

\begin{figure*}[h!]
\begin{center}
\includegraphics[width=\textwidth]{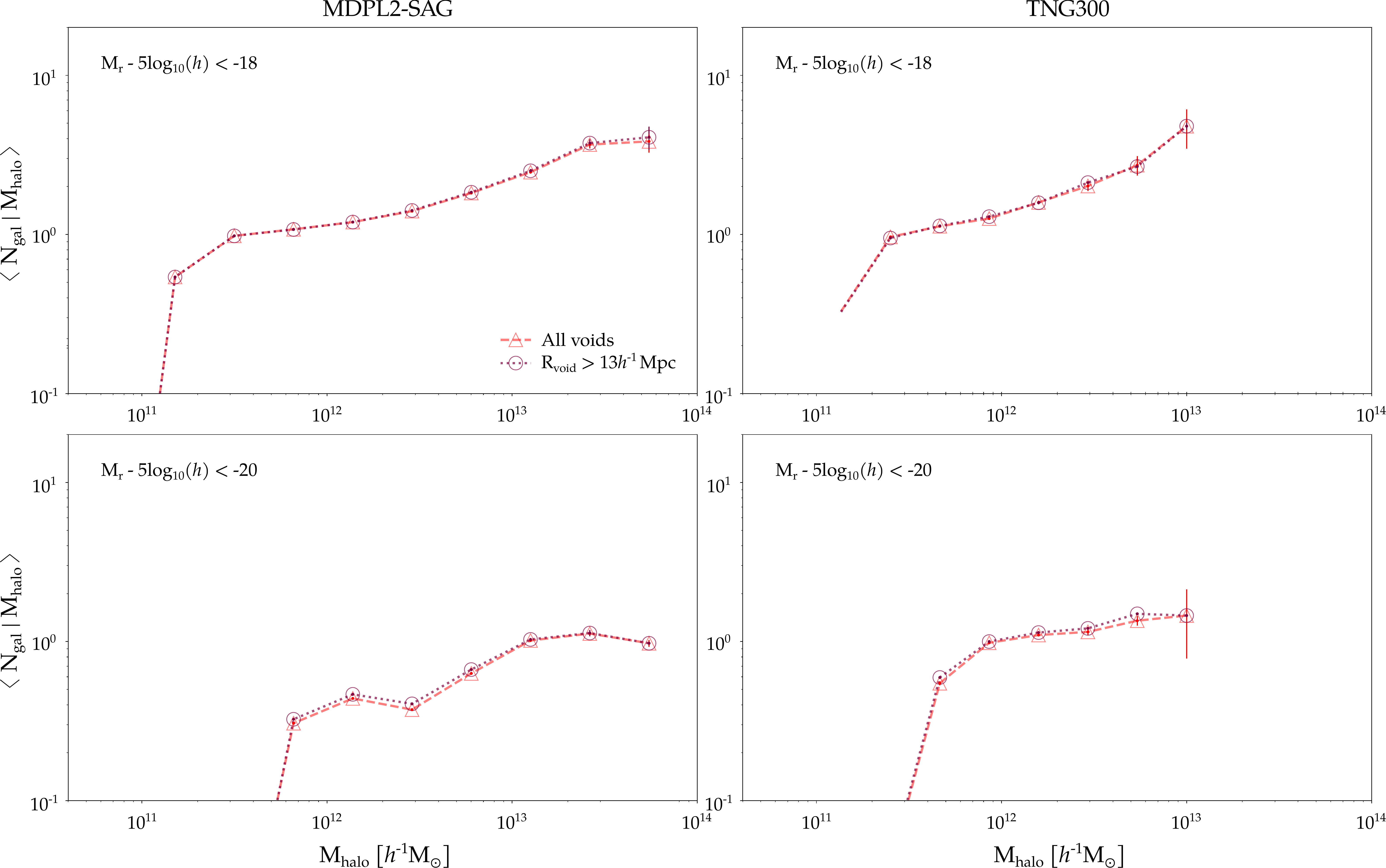}
\end{center}
\caption{\label{fig:HOD_Voids_CutRvoid} HOD in MDPL2-SAG and TNG300 catalogues for the complete samples of voids (triangles) and voids with $R_{\rm void} > 13h^{-1}$Mpc (circles). The left column shows the results for MDPL2-SAG and the right column for the TNG300. The top panels correspond to the limit magnitude $M_{\rm r} - 5\log_{\rm 10}(h)=-18$ and the bottom panels to the $M_{\rm r} - 5\log_{\rm 10}(h)=-20$. The uncertainties are calculated following the standard jackknife procedure.}
\end{figure*}

To analyze both the effects of differences in galaxy number density and the lack of completeness on our HOD measurements, we give in Fig. \ref{fig:HOD_Voids_CutRvoid} HOD measurements inside voids for the complete sample (circles) and that obtained considering 
only voids with $R_{\rm void} > 13h^{-1}$Mpc (triangles). We perform this comparison for TNG300 (right panels) and MDPL2-SAG (left panels) catalogues for two absolute magnitude thresholds, $M_{\rm r} - 5\log_{\rm 10}(h)=-18$ and $-20$ showin in the upper and lower panels, respectively. As it can be seen, by restricting  the number density of voids cutting by void radius does not produce a significant change
in the HOD determination. We conclude that both, the lack of completeness of small voids, and the differences in number density of voids between the catalogues, do not significantly affect the  results of our work.

\subsection{Equal number density of galaxies and voids}
\label{sec:ap2}

As mentioned above, we have identified voids and calculated the HOD using the same absolute magnitude thresholds. However, we can
perform this analysis with the same number density of tracers in order to obtain a similar number density distribution of voids.

\begin{figure}[h]
\begin{center}
\includegraphics[width=\columnwidth]{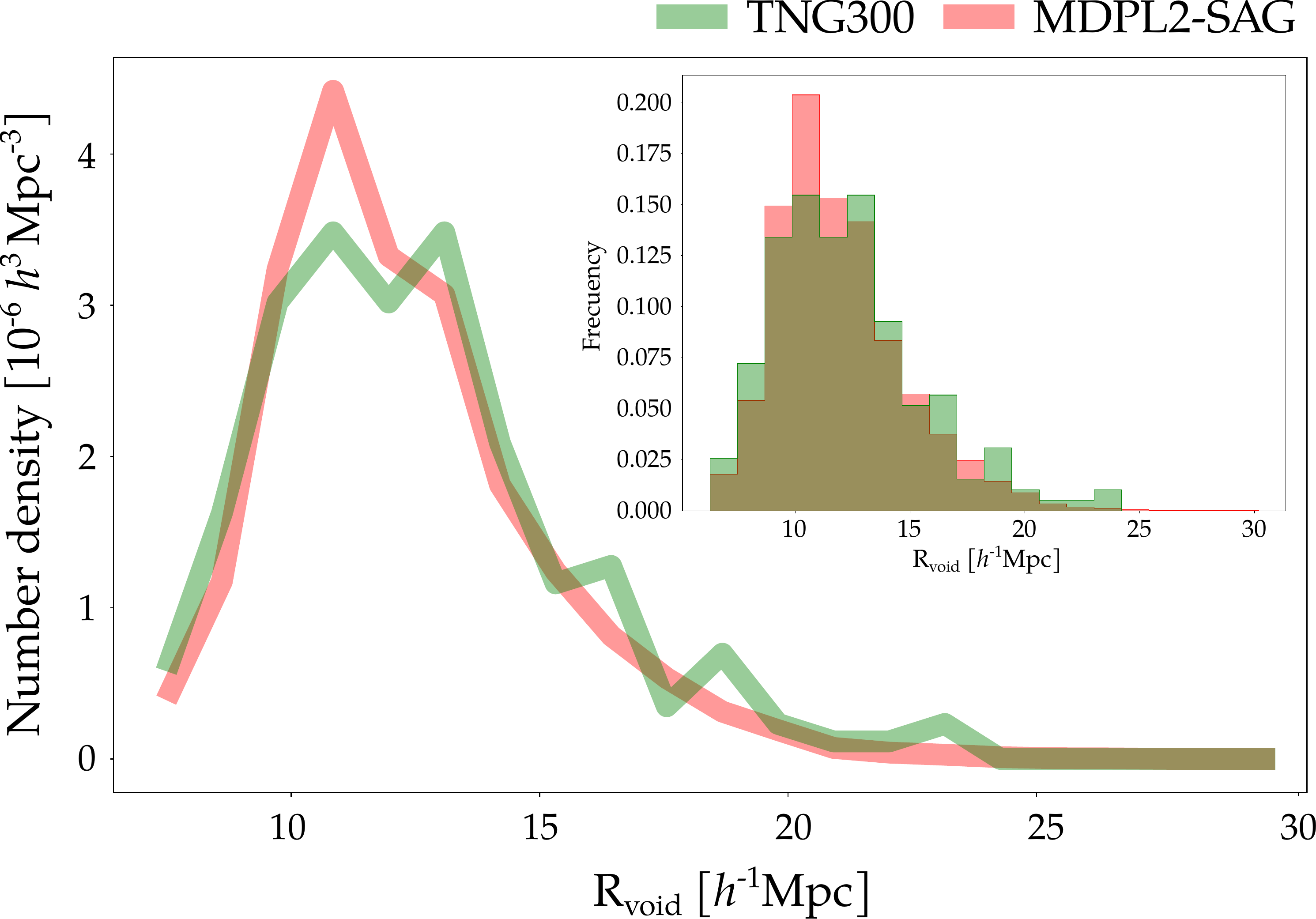}
\end{center}
\caption{\label{fig:VoidDensity_Idem} Same as Fig. \ref{fig:VoidDensity} but for voids identified using the same number density of tracers. Again, green and red curves correspond to TNG300 and MDPL2-SAG catalogues, respectively, and the inset show the size distributions of voids.}
\end{figure}

In Fig. \ref{fig:VoidDensity_Idem} we show the volume number density of voids as a function of radius when the void identification
is restricted to the same number density of tracers, $n_{\rm g} = 6.37 \times 10^{-3}~h^3{\rm Mpc}^{-3}$, for both catalogues. The green curve corresponds to TNG300 and the red to MDPL2-SAG. The inset in the figure shows the radii distribution for both void samples. 
As it can be seen, these results differ from those presented in Fig. \ref{fig:VoidDensity}. As expected, now both void catalogues show the same  volume number density and size distribution in the complete range of $R_{\rm void}$.

Using these new void catalogues, we use two cuts in number density of galaxies: $n_{\rm g} = 31.42 \times 10^{-3}h^{3}{\rm Mpc}^{-3}$ and $6.37 \times 10^{-3}h^{3}{\rm Mpc}^{-3}$, which correspond to the number density of galaxies for $M_{\rm r}=-18$ and $M_{\rm r}=-20$, respectively, from the Sloan Digital Sky Survey \citep{Guo2015}. 

\begin{figure*}
\begin{center}
\includegraphics[width=0.9\textwidth]{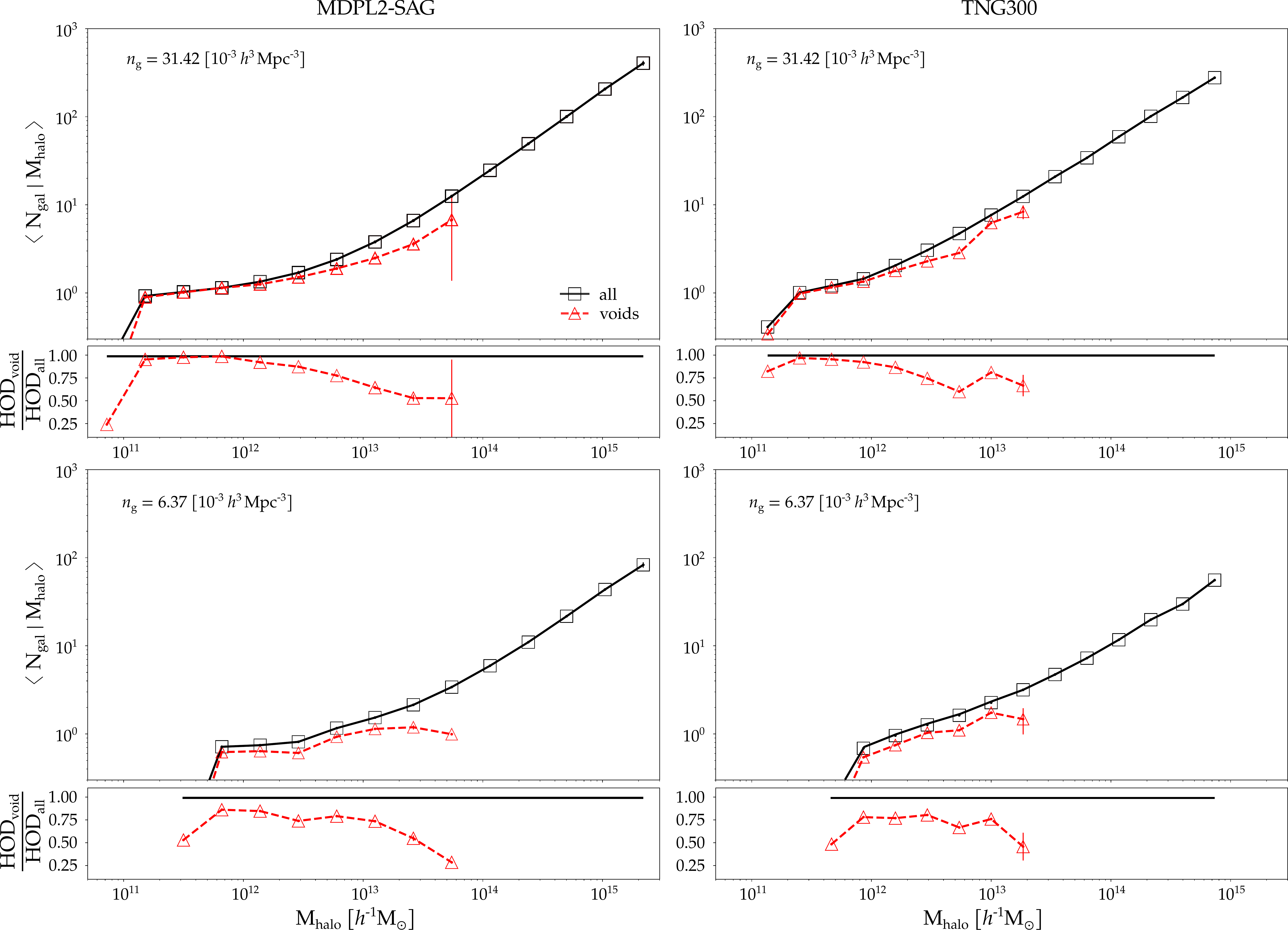}
\end{center}
\caption{\label{fig:HOD_Voids_CutDensity} HOD in MDPL2-SAG (left) and TNG300 (right) catalogues for two thresholds of number density of galaxies, $n_{\rm g}=31.42\times 10^{-3}~h^3{\rm Mpc}^{-3}$ (top panels) and $n_{\rm g}=6.37\times 10^{-3}~h^3{\rm Mpc}^{-3}$ (bottom panels). Black solid lines represent the overall HOD, meanwhile red dashed lines the HOD measured inside the voids. For each number density and catalogue, the ratio between both HODs are shown at the bottom of each panel. Quoted uncertainties were calculated through the standard jackknife procedure.}
\end{figure*}

Fig \ref{fig:HOD_Voids_CutDensity} shows the comparison between the HOD inside voids (red dashed lines) with the overall results (black solid lines) for MDPL2-SAG (left panels) and TNG300 (right panels) catalogues.
Small bottom panels show the ratio between HOD in voids and HOD in the complete sample. 
It is clear that the results are completely consistent with those obtained in the main part of this work, where we use fixed absolute magnitude cuts. It is worth to mention that the same behaviour is obtained for other equal number density cuts.
We believe that it is simpler to use a luminosity threshold to compare the simulations and observational data as well, 
so our main studies were performed under this prescription.

\end{appendix}

\end{document}